\documentclass[12pt]{article}
\usepackage{amsfonts,graphicx,amsmath,amsthm,amssymb,epsfig}
\usepackage{float}
\allowdisplaybreaks
\begin{document}

\title{\bf Complexity Analysis of Charged Dynamical Dissipative Cylindrical Structure in Modified Gravity}
\author{M. Sharif$^1$ \thanks{msharif.math@pu.edu.pk} and Tayyab Naseer$^{1,2}$ \thanks{tayyabnaseer48@yahoo.com}\\
$^1$ Department of Mathematics and Statistics, The University of Lahore,\\
1-KM Defence Road Lahore, Pakistan.\\
$^2$ Department of Mathematics, University of the Punjab,\\
Quaid-i-Azam Campus, Lahore-54590, Pakistan.}

\date{}
\maketitle

\begin{abstract}
This article focuses on the formulation of some scalar factors which
are uniquely expressed in terms of matter variables for dynamical
charged dissipative cylindrical geometry in a standard gravity model
$\mathcal{R}+\Phi\mathcal{Q}$ ($\Phi$ is the coupling parameter,
$\mathcal{Q}=\mathcal{R}_{\varphi\vartheta}\mathcal{T}^{\varphi\vartheta}$)
and calculates four scalars by orthogonally decomposing the Riemann
tensor. We find that only $\mathcal{Y}_{TF}$ involves inhomogeneous
energy density, heat flux, charge and pressure anisotropy coupled
with modified corrections, and thus call it as complexity factor for
the considered distribution. Two evolutionary modes are discussed to
study the dynamics of cylinder. We then take the homologous
condition with $\mathcal{Y}_{TF}=0$ to calculate unknown metric
potentials in the absence as well as presence of heat dissipation.
The stability criterion of the later condition is also checked
throughout the evolution by applying some constraints. We conclude
that the effects of charge and modified theory yield more complex
system.
\end{abstract}
{\bf Keywords:}
$f(\mathcal{R},\mathcal{T},\mathcal{R}_{\varphi\vartheta}\mathcal{T}^{\varphi\vartheta})$
gravity; Self-gravitating systems; Complexity factor. \\
{\bf PACS:} 04.40.-b; 04.40.Dg; 04.50.Kd.

\section{Introduction}

The accelerating expansion of our universe has recently been viewed
from several remarkable cosmic observations such as redshift and
distance-luminosity relationship of type IA Supernovae \cite{1a,1b}.
The study of current nature of the universe within the context of
general relativity ($\mathbb{GR}$) suffers from some shortcomings
like fine-tuning and cosmic coincidence. In view of this,
researchers modified $\mathbb{GR}$ to find a suitable solution to
such issues and include the effects of rapid expansion. The
immediate generalization of $\mathbb{GR}$ was proposed to study
cosmological consequences at large scale, named as $f(\mathcal{R})$
theory \cite{1c}. The Einstein-Hilbert action was modified by
replacing the Ricci scalar with its generic function
$f(\mathcal{R})$ to get the effects of this extended theory.
Numerous astrophysicists \cite{2}-\cite{2d} studied different models
in this framework and obtained physically feasible compact
structures by employing multiple approaches.

Bertolami et al. \cite{10} presented the notion of coupling between
matter and geometry in $f(\mathcal{R})$ gravity for the very first
time by engaging the effects of geometry in matter Lagrangian
through insertion of the Ricci scalar. A couple of years ago,
multiple extensions of $\mathbb{GR}$ encompassing such interaction
were proposed that prompted astronomers to explore the physical
feasibility of modified gravitational models. This idea has recently
been generalized at action level by Harko et al. \cite{20} by
introducing $f(\mathcal{R},\mathcal{T})$ theory, in which
$\mathcal{T}$ is trace of the energy-momentum tensor
$(\mathbb{EMT})$. The contribution of the $\mathbb{EMT}$ in analytic
functional of any theory results in its non-zero divergence contrary
to $\mathbb{GR}$ and $f(\mathcal{R})$ framework. A large body of
literature \cite{21}-\cite{21f} exists to analyze the effects of
coupling on self-gravitating structures and found several remarkable
results in this theory. Haghani et al. \cite{22} generalized this
gravity by adding a factor $\mathcal{Q}$ which is the contraction of
the Ricci tensor and $\mathbb{EMT}$. This theory explains the
inflationary era of our cosmos properly. They also studied some
cosmological applications corresponding to three different
$f(\mathcal{R},\mathcal{T},\mathcal{Q})$ models like
$\mathcal{R}+\lambda\mathcal{Q},~\mathcal{R}(1+\lambda\mathcal{Q})$
and $\mathcal{R}+\zeta\sqrt {|\mathcal{T}|}+\lambda\mathcal{Q}$.

By adopting first two of the above models along with matter
Lagrangian as $\mathbb{L}_m=\rho,~-P$, Sharif and Zubair \cite{22a}
discussed thermodynamical laws of black hole and computed some
acceptable values of the coupling parameter $\lambda$. They also
generalized energy bounds for this scenario and obtained some
constraints for which those bounds show viable behavior \cite{22b}.
They found that energy conditions do not hold for negative values of
$\lambda$. Odintsov and S\'{a}ez-G\'{o}mez \cite{23} discussed
several cosmological solutions in
$f(\mathcal{R},\mathcal{T},\mathcal{Q})$ gravity by reconstructing
their corresponding gravitational action. Sharif and Waseem
\cite{25a} analyzed three different compact stars in this context
along with matter Lagrangian as $\mathbb{L}_m=-P_r,~-P_\bot$. They
concluded that these systems show stable behavior near the center
for $\mathbb{L}_m=-P_r$ only. Yousaf et al. \cite{26,26a} computed
modified structure scalars for effective $\mathbb{EMT}$ with and
without charge in spherical system and discussed the evolution of
non-static self-gravitating structures. The complexity of
self-gravitating systems has also been measured through a scalar
$\mathcal{Y}_{TF}$ \cite{26d,26e}. We have obtained some stable
anisotropic solutions by employing multiple approaches in this
context \cite{27,27aa}.

The self-gravitating structures whose interior is cylindrically
symmetric have been supported by the existence of cylindrical
gravitational waves. The study of such geometrical objects produces
significant consequences, and thus motivated many astrophysicists to
investigate their fundamental features. The pioneering study of
these massive systems have been done by Bronnikov and Kovalchuk
\cite{35d}. Wang \cite{35e} determined analytic solutions to the
field equations corresponding to four-dimensional cylindrical
geometry along with a massless scalar field. They further observed
the formation of a black hole as a result of collapse of such body.
The influence of electromagnetic field on massive objects plays a
considerable role in studying their evolution and stability. The
attractive nature of gravity can be overcome through the magnetic as
well as Coulomb forces. A huge amount of charge is needed to hamper
the gravitational attraction and sustain the stable behavior of
self-gravitating systems. Bekenstein \cite{27a} investigated
collapse of charged spherical structure and found that charge
reduces the collapse rate. The same result has been produced by
Esculpi and Aloma \cite{27b} while studying collapsing phenomenon
for anisotropic charged distribution. Sharif and Azam \cite{27c}
examined the impact of charge on the evolution of cylindrically
symmetric system. Takisa an Maharaj \cite{27d} discussed anisotropic
gravitating body by considering polytropic equation of state and
found the profiles of matter variables which are consistent with
earlier treatments.

Numerous massive and highly dense structures (stars and galaxies)
are the main constituents that made the visible portion of our
universe. These self-gravitating systems incorporate different
physical quantities such as energy density, pressure, heat flow in
their interiors that may cause to make them complex. A mathematical
definition of structural complexity is required in terms of physical
factors. In this regard, L{\'o}pez-Ruiz et al. \cite{28} defined
complexity for the very first time in terms of information and
entropy. This definition was initially employed on two simplest
physical systems (perfect crystal and ideal gas). The molecules in
former structure are symmetrically arranged throughout and thus has
zero entropy whereas it is maximum in ideal gas as particles are
randomly distributed. Moreover, ideal gas and perfect crystal
contain maximum and less data (or information) in accordance with
their structural composition, respectively. However, both patterns
have no complexity.

Later, this concept was proposed in terms of disequilibrium but
failed because the complexity of both the structures has been found
to be zero under this definition \cite{29,29a}. Another definition
was suggested through energy density that replaced the probability
distribution \cite{30,30a}, nonetheless this was insufficient as the
interior of compact geometry may involve some other variables (heat
flux, pressure and temperature, etc). Herrera \cite{31} recently
redefined this concept and stated that complexity can be measured in
terms of energy density inhomogeneity and pressure anisotropy inside
a static sphere. He named a particular scalar as the complexity
factor (encompassing all aforesaid parameters) that comes from
orthogonal decomposition of the Riemann tensor. Sharif and Butt
\cite{32,32a} analyzed the effects of charge on this factor and also
studied for uncharged cylindrical fluid source. Herrera et al.
\cite{33} then studied a dynamical dissipative system and discussed
some evolutionary patterns along with kinematical/dynamical
quantities. This work has also been generalized to the axially
symmetric spacetime \cite{34}. Sharif and Majid \cite{35}-\cite{35c}
found several solutions for self-gravitating systems by extending
this definition in Brans-Dicke scenario. The complexity for
anisotropic configurations has also been analyzed in the context of
$f(\mathcal{R},\mathcal{T})$ and $f(\mathcal{G},\mathcal{T})$
theories \cite{35ca}-\cite{36b}.

This article addresses evolution and complex composition of the
charged dynamical cylinder involving the effects of heat dissipation
in
$f(\mathcal{R},\mathcal{T},\mathcal{R}_{\varphi\vartheta}\mathcal{T}^{\varphi\vartheta})$
gravity. The paper is outlined as follows. We introduce basic
formalism of this modified theory and calculate the corresponding
field equations as well as Bianchi identities for the model
$\mathcal{R}+\zeta\mathcal{R}_{\varphi\vartheta}\mathcal{T}^{\varphi\vartheta}$
in section \textbf{2}. Section \textbf{3} discusses four structure
scalars that come from orthogonal splitting of the Riemann tensor.
We further study the evolution of considered matter source through
some evolutionary modes in section \textbf{4}. The unknown metric
potentials of cylindrical geometry are determined in the
absence/presence of heat dissipation in section \textbf{5}. Section
\textbf{6} explores some conditions which may deviate the system
from complexity-free scenario. Section \textbf{7} summarizes all our
findings.

\section{The $f(\mathcal{R},\mathcal{T},\mathcal{R}_{\varphi\vartheta}\mathcal{T}^{\varphi\vartheta})$ Gravity}

The generic function of $\mathcal{R},~\mathcal{T}$ and $Q$ in place
of the Ricci scalar in the Einstein-Hilbert action (with
$\kappa=8\pi$) provides the following form \cite{23}
\begin{equation}\label{g1}
\mathbb{S}_{f(\mathcal{R},\mathcal{T},\mathcal{R}_{\varphi\vartheta}\mathcal{T}^{\varphi\vartheta})}=\int\sqrt{-g}
\left\{\frac{f(\mathcal{R},\mathcal{T},\mathcal{R}_{\varphi\vartheta}\mathcal{T}^{\varphi\vartheta})}{16\pi}
+\mathbb{L}_{\mathcal{M}}+\mathbb{L}_{\mathcal{EM}}\right\}d^{4}x,
\end{equation}
where $\mathbb{L}_{\mathcal{M}}$ and $\mathbb{L}_{\mathcal{EM}}$ are
the Lagrangian densities corresponding to matter distribution and
electromagnetic field, respectively. The execution of the
variational principle on the action \eqref{g1} yields the field
equations as
\begin{equation}\label{g2}
\mathcal{G}_{\varphi\vartheta}=\mathcal{T}_{\varphi\vartheta}^{(\mathrm{EFF})}=\frac{8\pi}{f_{\mathcal{R}}-\mathbb{L}_{\mathcal{M}}
f_{\mathcal{Q}}}\big(\mathcal{T}_{\varphi\vartheta}+\mathcal{E}_{\varphi\vartheta}\big)+\mathcal{T}_{\varphi\vartheta}^{(\mathcal{C})}.
\end{equation}
Here, $\mathcal{G}_{\varphi\vartheta}$ is the Einstein tensor,
$\mathcal{T}_{\varphi\vartheta}^{(\mathrm{EFF})}$ is termed as the
$\mathbb{EMT}$ in modified framework,
$\mathcal{T}_{\varphi\vartheta}$ is the anisotropic matter
$\mathbb{EMT}$ and $\mathcal{E}_{\varphi\vartheta}$ is the
electromagnetic tensor. The last factor in the above equation has
the form
\begin{eqnarray}\nonumber
\mathcal{T}_{\varphi\vartheta}^{(\mathcal{C})}&=&-\frac{1}{\big(\mathbb{L}_{\mathcal{M}}f_{\mathcal{Q}}-f_{\mathcal{R}}\big)}
\left[\left(f_{\mathcal{T}}+\frac{1}{2}\mathcal{R}f_{\mathcal{Q}}\right)\mathcal{T}_{\varphi\vartheta}
+\left\{\frac{\mathcal{R}}{2}(\frac{f}{\mathcal{R}}-f_{\mathcal{R}})-\mathbb{L}_{\mathcal{M}}f_{\mathcal{T}}\right.\right.\\\nonumber
&-&\left.\frac{1}{2}\nabla_{\varrho}\nabla_{\omega}(f_{\mathcal{Q}}\mathcal{T}^{\varrho\omega})\right\}g_{\varphi\vartheta}
-\frac{1}{2}\Box(f_{\mathcal{Q}}\mathcal{T}_{\varphi\vartheta})-(g_{\varphi\vartheta}\Box-
\nabla_{\varphi}\nabla_{\vartheta})f_{\mathcal{R}}\\\label{g4}
&-&2f_{\mathcal{Q}}\mathcal{R}_{\varrho(\varphi}\mathcal{T}_{\vartheta)}^{\varrho}
+\nabla_{\varrho}\nabla_{(\varphi}[\mathcal{T}_{\vartheta)}^{\varrho}
f_{\mathcal{Q}}]+2(f_{\mathcal{Q}}\mathcal{R}^{\varrho\omega}+\left.f_{\mathcal{T}}g^{\varrho\omega})\frac{\partial^2\mathbb{L}_{\mathcal{M}}}
{\partial g^{\varphi\vartheta}\partial g^{\varrho\omega}}\right].
\end{eqnarray}
The partial differentiation of $f$ is
$f_{\mathcal{R}}=\frac{\partial
f(\mathcal{R},\mathcal{T},\mathcal{Q})}{\partial
\mathcal{R}}$,~$f_{\mathcal{T}}=\frac{\partial
f(\mathcal{R},\mathcal{T},\mathcal{Q})}{\partial \mathcal{T}}$ and
$f_{\mathcal{Q}}=\frac{\partial
f(\mathcal{R},\mathcal{T},\mathcal{Q})}{\partial \mathcal{Q}}$.
Also, $\nabla_\varrho$ is the covariant derivative and $\Box$ is the
D'Alambert operator whose mathematical expression is $\Box\equiv
\frac{1}{\sqrt{-g}}\partial_\varphi\big(\sqrt{-g}g^{\varphi\vartheta}\partial_{\vartheta}\big)$.
The most suitable choice of the matter Lagrangian in this case is
$\mathbb{L}_{\mathcal{M}}=-\frac{1}{4}\mathcal{N}_{\varphi\vartheta}\mathcal{N}^{\varphi\vartheta}$
which results in $\frac{\partial^2\mathbb{L}_{\mathcal{M}}}
{\partial g^{\varphi\vartheta}\partial
g^{\varrho\omega}}=-\frac{1}{2}\mathcal{N}_{\varphi\varrho}\mathcal{N}_{\vartheta\omega}$
\cite{22}. Here,
$\mathcal{N}_{\varphi\vartheta}=\mathcal{O}_{\vartheta;\varphi}-\mathcal{O}_{\varphi;\vartheta}$
is known as the Maxwell field tensor and $\mathcal{O}_{\vartheta}$
serves as the four potential.

The $\mathbb{EMT}$ describing anisotropic configuration with heat
dissipation is
\begin{equation}\label{g5}
\mathcal{T}_{\varphi\vartheta}=\mu \mathcal{K}_{\varphi}
\mathcal{K}_{\vartheta}+P
h_{\varphi\vartheta}+\Pi_{\varphi\vartheta}+\varsigma\big(\mathcal{K}_\varphi\mathcal{W}_\vartheta
+\mathcal{W}_\varphi\mathcal{K}_\vartheta\big),
\end{equation}
where
$\mathcal{W}_{\varphi},~\mathcal{K}_\varphi~h_{\varphi\vartheta}$
and $\varsigma$ are the four-vector, four-velocity, projection tenor
and heat flux, respectively which satisfy the relations
$\mathcal{K}_{\varphi}
\mathcal{K}^{\varphi}=-1,~\mathcal{K}_{\varphi}
\mathcal{W}^{\varphi}=0,~\varsigma_{\varphi}\mathcal{K}^{\varphi}=0,~\mathcal{W}_{\varphi}
\mathcal{W}^{\varphi}=1$. The remaining terms are described as
\begin{align}\label{g5b}
P&=\frac{P_r+2P_\bot}{3}, \quad
h_{\varphi\vartheta}=g_{\varphi\vartheta}+\mathcal{K}_{\varphi}
\mathcal{K}_{\vartheta},\\\label{g5c}
\Pi_{\varphi\vartheta}&=\Pi\bigg(\mathcal{W}_{\varphi}\mathcal{W}_{\vartheta}-\frac{h_{\varphi\vartheta}}{3}\bigg),
\quad \Pi=P_r-P_\bot.
\end{align}
It is important to mention here that pressure generally exists in
three different directions for anisotropic cylindrically symmetric
structure \cite{36c,36d}, but the matter content \eqref{g5} is not
the most general form of the fluid distribution, rather it is a
restricted case.

As this theory involves components of fluid configuration coupled
with geometry, thus the equivalence principle does not hold and
divergence of the corresponding $\mathbb{EMT}$ is no more conserved,
i.e., $\nabla_\varphi \mathcal{T}^{\varphi\vartheta}\neq 0$. This
causes the exertion of an extra force due to which test particles in
the gravitational field start their motion in non-geodesic path.
Consequently, we have
\begin{align}\nonumber
\nabla^\varphi
\big(\mathcal{T}_{\varphi\vartheta}+\mathcal{E}_{\varphi\vartheta})&=\frac{2}{2f_\mathcal{T}+\mathcal{R}f_\mathcal{Q}+16\pi}\bigg[\nabla_\varphi
\big(f_\mathcal{Q}\mathcal{R}^{\varrho\varphi}\mathcal{T}_{\varrho\vartheta}\big)
+\nabla_\vartheta\big(\mathbb{L}_\mathcal{M}f_\mathcal{T}\big)\\\nonumber
&-\mathcal{G}_{\varphi\vartheta}\nabla^\varphi\big(f_\mathcal{Q}\mathbb{L}_\mathcal{M}\big)-\frac{1}{2}\nabla_\vartheta\mathcal{T}^{\varrho\omega}
\big(f_\mathcal{T}g_{\varrho\omega}+f_\mathcal{Q}\mathcal{R}_{\varrho\omega}\big)\\\label{g4a}
&-\frac{1}{2}\big\{\nabla^{\varphi}(\mathcal{R}f_{\mathcal{Q}})
+2\nabla^{\varphi}f_{\mathcal{T}}\big\}\mathcal{T}_{\varphi\vartheta}\bigg].
\end{align}
The trace of $f(\mathcal{R},\mathcal{T},\mathcal{Q})$ field
equations yields
\begin{align}\nonumber
&3\nabla^{\varrho}\nabla_{\varrho}
f_\mathcal{R}-\mathcal{R}\left(\frac{\mathcal{T}}{2}f_\mathcal{Q}-f_\mathcal{R}\right)-\mathcal{T}(8\pi+f_\mathcal{T})+\frac{1}{2}
\nabla^{\varrho}\nabla_{\varrho}(f_\mathcal{Q}\mathcal{T})\\\nonumber
&+\nabla_\varphi\nabla_\varrho(f_\mathcal{Q}\mathcal{T}^{\varphi\varrho})-2f+(\mathcal{R}f_\mathcal{Q}+4f_\mathcal{T})\mathbb{L}_\mathcal{M}
+2\mathcal{R}_{\varphi\varrho}\mathcal{T}^{\varphi\varrho}f_\mathcal{Q}\\\nonumber
&-2g^{\vartheta\xi} \frac{\partial^2\mathbb{L}_\mathcal{M}}{\partial
g^{\vartheta\xi}\partial
g^{\varphi\varrho}}\left(f_\mathcal{T}g^{\varphi\varrho}+f_\mathcal{Q}\mathcal{R}^{\varphi\varrho}\right)=0.
\end{align}
The gravitational effects of $f(\mathcal{R},\mathcal{T})$ theory can
be obtained by considering $f_\mathcal{Q}=0$, whereas
$f_\mathcal{T}=0$ with previous limit gives $f(\mathcal{R})$
gravity. The electromagnetic tensor is
\begin{equation*}
\mathcal{E}_{\varphi\vartheta}=\frac{1}{4\pi}\left[\frac{1}{4}g_{\varphi\vartheta}\mathcal{N}^{\rho\omega}\mathcal{N}_{\rho\omega}
-\mathcal{N}^{\omega}_{\varphi}\mathcal{N}_{\omega\vartheta}\right],
\end{equation*}
and Maxwell equations have the form
\begin{equation*}
\mathcal{N}^{\varphi\vartheta}_{;\vartheta}=4\pi
\mathcal{J}^{\varphi}, \quad
\mathcal{N}_{[\varphi\vartheta;\omega]}=0,
\end{equation*}
where $\mathcal{J}^{\varphi}=\varpi \mathcal{K}^{\varphi}$. Here,
$\mathcal{J}^{\varphi}$ and $\varpi$ indicate the current and charge
densities, respectively.

We adopt a standard model of the form (proposed by Haghani et al.
\cite{22})
\begin{equation}\label{g5d}
f(\mathcal{R},\mathcal{T},\mathcal{R}_{\varphi\vartheta}\mathcal{T}^{\varphi\vartheta})=f_1(\mathcal{R})+f_2
(\mathcal{R}_{\varphi\vartheta}\mathcal{T}^{\varphi\vartheta})=\mathcal{R}+\Phi
\mathcal{R}_{\varphi\vartheta}\mathcal{T}^{\varphi\vartheta}.
\end{equation}
This model offers several solutions showing an oscillatory behavior
for positive values of $\Phi$, while $\Phi < 0$ produces the cosmic
scale factor which has a hyperbolic cosine-type dependence. It is
worth noting that the value of the coupling parameter within its
observed range guarantees physical feasibility of the corresponding
gravity model. Some acceptable values of $\Phi$ have been explored
under which the obtained solutions with respect to the model
\eqref{g5d} for isotropic configuration show stable behavior
\cite{22a,22b}.

We take restricted form of cylindrically symmetric dynamical line
element to examine the interior as
\begin{equation}\label{g6}
ds^2=-A^2 dt^2+B^2 dr^2+C^2(d\theta^2+\alpha^2 dz^2),
\end{equation}
where $A=A(t,r)$ and $B=B(t,r)$ are dimensionless, while $C=C(t,r)$
has the dimension as of $r$. Also, $\alpha$ is supposed to be a
constant with dimension of inverse length. The charge inside the
cylinder is defined as
\begin{equation}\label{g6a}
s(r)=4\pi\int_{0}^{r}\varpi BC^2dr,
\end{equation}
and the matter Lagrangian becomes
$\mathbb{L}_{\mathcal{M}}=\frac{s^2}{2C^4}$. The four-velocity, heat
flux and four-vector in comoving scenario are characterized as
\begin{equation}\label{g7}
\mathcal{K}^\varphi=\delta^\varphi_0 A^{-1}, \quad
\varsigma^\varphi=\delta^\varphi_1 \varsigma B^{-1}, \quad
\mathcal{W}^\varphi=\delta^\varphi_1 B^{-1}.
\end{equation}
The quantity $\mathcal{Q}$ of the model \eqref{g5d} becomes
\begin{align}\nonumber
\mathcal{Q}&=-\frac{1}{A^3B^3C}\bigg[\mu\big\{2\ddot{C}AB^3+\ddot{B}AB^2C-A''A^2BC-2\dot{A}\dot{C}B^3-2A'C'A^2B\\\nonumber
&+A'B'A^2C-\dot{A}\dot{B}B^2C\big\}+4\varsigma
AB\big\{\dot{C}'AB-A'\dot{C}B-\dot{B}C'A\big\}\\\nonumber
&+P_r\big\{2C''A^3B-\ddot{B}AB^2C+A''A^2BC-2\dot{B}\dot{C}AB^2-2B'C'A^3\\\nonumber
&-A'B'A^2C+\dot{A}\dot{B}B^2C\big\}+\frac{2P_\bot}{C}\big\{C''A^3BC-\ddot{C}AB^3C-\dot{C}^2AB^3\\\nonumber
&-\dot{B}\dot{C}AB^2C+\dot{A}\dot{C}CB^3+C'^2A^3B-B'C'A^3C+A'C'A^2BC\big\}\bigg],
\end{align}
where $.=\frac{\partial}{\partial t}$ and
$'=\frac{\partial}{\partial r}$. The non-vanishing components of the
field equations \eqref{g2} are
\begin{align}\nonumber
8\pi\big(\bar{\mu}+\frac{\bar{s}^2}{8\pi
C^4}+\mathcal{T}^{0(\mathcal{C})}_{0}+\mathcal{E}^{0(\mathcal{C})}_{0}\big)&=
-\frac{1}{B^2}\bigg(\frac{C'^2}{C^2}+\frac{2C''}{C}-\frac{2C'B'}{CB}\bigg)\\\label{g8}
&+\frac{\dot{C}}{CA^2}\bigg(\frac{\dot{C}}{C}+\frac{2\dot{B}}{B}\bigg),\\\label{g8a}
8\pi\big(-\bar{\varsigma}+\mathcal{T}^{1(\mathcal{C})}_{0}+\mathcal{E}^{1(\mathcal{C})}_{0}\big)&=\frac{1}{AB}\bigg(\frac{2A'\dot
C}{A C}+\frac{2C'\dot B}{C B}-\frac{2\dot C'}{C}\bigg), \\\nonumber
8\pi\big(\bar{P_r}-\frac{\bar{s}^2}{8\pi
C^4}+\mathcal{T}^{1(\mathcal{C})}_{1}+\mathcal{E}^{1(\mathcal{C})}_{1}\big)&=
-\frac{1}{A^2}\bigg\{\frac{2\ddot{C}}{C}-\frac{\dot{C}}{C}\bigg(\frac{2\dot{A}}{A}-\frac{\dot{C}}{C}\bigg)\bigg\}\\\label{g8b}
&+\frac{1}{B^2}\bigg(\frac{C'^2}{C^2}+\frac{2A'C'}{AC}\bigg),\\\nonumber
8\pi\big(\bar{P_\bot}+\frac{\bar{s}^2}{8\pi
C^4}+\mathcal{T}^{2(\mathcal{C})}_{2}+\mathcal{E}^{2(\mathcal{C})}_{2}\big)&=
-\frac{1}{A^2}\bigg\{\frac{\dot{B}\dot{C}}{BC}+\frac{\ddot
B}{B}+\frac{\ddot C}{C}-\frac{\dot A}{A}\bigg(\frac{\dot
C}{C}+\frac{\dot B}{B}\bigg)\bigg\}\\\label{g8c}
&+\frac{1}{B^2}\bigg\{\bigg(\frac{A'}{A}-\frac{B'}{B}\bigg)\frac{C'}{C}-\frac{A'B'}{AB}+\frac{A''}{A}+\frac{C''}{C}\bigg\},
\end{align}
where $\bar{\mu}=\frac{\mu}{1-\frac{\Phi
s^2}{2C^4}},~\bar{\varsigma}=\frac{\varsigma}{1-\frac{\Phi
s^2}{2C^4}},~\bar{P_r}=\frac{P_{r}}{1-\frac{\Phi
s^2}{2C^4}},~\bar{P_\bot}=\frac{P_\bot}{1-\frac{\Phi s^2}{2C^4}}$
and $\bar{s}^2=\frac{s^2}{1-\frac{\Phi s^2}{2C^4}}$. The terms
$\mathcal{T}^{0(\mathcal{C})}_{0},~\mathcal{T}^{1(\mathcal{C})}_{0},~\mathcal{T}^{1(\mathcal{C})}_{1}$
and $\mathcal{T}^{2(\mathcal{C})}_{2}$ as well as their
corresponding charge components
$\mathcal{E}^{0(\mathcal{C})}_{0},~\mathcal{E}^{1(\mathcal{C})}_{0},~\mathcal{E}^{1(\mathcal{C})}_{1}$
and $\mathcal{E}^{2(\mathcal{C})}_{2}$ are modified corrections to
the field equations whose values are given in Appendix $\mathbf{A}$.

The non-null components of Bianchi identity through Eq.\eqref{g4a}
are given as
\begin{align}\nonumber
\mathcal{T}^{\varphi\vartheta}_{\quad;\vartheta}\mathcal{K}_{\varphi}&=-\frac{1}{A}\bigg\{\dot{\mu}+\big(\mu+P_r\big)\frac{\dot{B}}{B}
+2\big(\mu+P_\bot\big)\frac{\dot{C}}{C}\bigg\}-\frac{1}{B}\bigg\{\varsigma'+2\varsigma\bigg(\frac{A'}{A}+\frac{C'}{C}\bigg)\bigg\}\\\label{g11}
&=\mathbb{Z}_1-\frac{\zeta}{A(16\pi+\zeta\mathcal{R})}\bigg(\frac{\mu\dot{\mathcal{R}}}{A}+\frac{\varsigma\mathcal{R}'}{B}\bigg),
\end{align}
and
\begin{align}\nonumber
\mathcal{T}^{\varphi\vartheta}_{\quad;\vartheta}\mathcal{W}_{\varphi}&=\frac{1}{A}\bigg\{2\varsigma\bigg(\frac{B'}{B}
+\frac{C'}{C}\bigg)+\dot{\varsigma}\bigg\}+\frac{1}{B}\bigg\{P_r'+2\big(P_r-P_\bot\big)\frac{C'}{C}+\big(\mu+P_r\big)\frac{A'}{A}\\\label{g11a}
&-\frac{ss'}{4\pi
C^4}\bigg\}=\mathbb{Z}_2-\frac{\zeta}{B(16\pi+\zeta\mathcal{R})}\bigg(\frac{P_r\mathcal{R}'}{B}+\frac{\varsigma\dot{\mathcal{R}}}{A}\bigg),
\end{align}
where the terms on the right hand side of the above equations
confirm the non-conservation of this gravity. The values of
$\mathbb{Z}_1$ and $\mathbb{Z}_2$ are included in Appendix
\textbf{A}. Some dynamical terms such as expansion scalar, non-zero
components of shear tensor as well as four-acceleration are defined
as
\begin{align}\label{g11b}
&\Theta=\frac{1}{A}\left(2\frac{\dot C}{C}+\frac{\dot
B}{B}\right),\\\label{g11c} &\sigma_{11}=\frac{2}{3}B^2\sigma, \quad
\sigma_{22}=\frac{\sigma_{33}}{\alpha^2}=-\frac{1}{3}C^2\sigma,\\\label{g11d}
&\sigma^{\varphi\vartheta}\sigma_{\varphi\vartheta}=\frac{2}{3}\sigma^2,~\sigma=\frac{1}{A}\left(\frac{\dot
B}{B}-\frac{\dot C}{C}\right),
\\\label{g11e} &a_{1}=\frac{A'}{A},\quad
a=\sqrt{a^{\varphi}a_{\varphi}}=\frac{A'}{AB}.
\end{align}
It should be noted that two scalars are needed to determine the
shear tensor in the case of general cylindrically symmetric fluid
\cite{36d}, however, we define a single scalar function \eqref{g11d}
due to restricted class of the cylindrical spacetime.

The effect of expansion scalar and shear on the fluid distribution
can be analyzed by the alternative form of Eq.\eqref{g8a} as
\begin{equation}\label{g12}
4\pi\big(\bar{\varsigma}-\mathcal{T}^{1(\mathcal{C})}_{0}-\mathcal{E}^{1(\mathcal{C})}_{0}\big)=\frac{1}{3}\left(\Theta-\sigma\right)'-\sigma
\frac{C'}{C}=\frac{C'}{B}\left[\frac{1}{3}\mathbb{D}_{\mathbb{R}}\left(\Theta-\sigma\right)-\frac{\sigma}{C}\right],
\end{equation}
where the proper radial derivative is symbolized as
$\mathbb{D}_{\mathbb{R}}=\frac{1}{C'}\frac{\partial}{\partial r}$.
The formula of C-energy can be employed to calculate mass of the
cylindrical geometry \cite{41ba}. Thus the following equation
provides relation between C-energy and the mass function as
\begin{equation}\label{g13}
\tilde{m}(t,r)=\mathfrak{L}\hat{\mathrm{E}}=\frac{\mathfrak{L}}{8}(1-\mathfrak{L}^{-2}\nabla_\varphi
\hat{r}\nabla^\varphi \hat{r}),
\end{equation}
where $\hat r=\varrho\mathfrak{L}$, $\varrho$ is the circumference
radius and $\mathfrak{L}$ symbolizes specific length.
Mathematically, we have
$\varrho^2=\eta_{(1)\vartheta}\eta_{(1)}^{\vartheta}$ and
$\mathfrak{L}^2=\eta_{(2)\vartheta}\eta_{(2)}^{\vartheta}$ in which
$\eta_{(1)}=\frac{\partial}{\partial \theta}$,
$\eta_{(2)}=\frac{\partial}{\partial z}$. Here, $\hat {\mathrm{E}}$
is defined as the gravitational energy per specific length. Equation
\eqref{g13} yields the mass in terms of metric potentials as
\begin{equation}\label{g14}
\tilde{m}=\frac{C}{2}\bigg[\frac{1}{4}-\bigg(\frac{C'}{B}\bigg)^2+\bigg(\frac{\dot
C}{A}\bigg)^2+\frac{s^2}{C^2}\bigg].
\end{equation}

We study the evolution of charged dynamical cylinder by utilizing
the definition of proper time derivative, i.e.,
$\mathbb{D}_{\mathbb{T}}=\frac{1}{A}\frac{\partial}{\partial t}$.
During the collapse of an astronomical object, the continuous
reduction of its radius occurs, as a result of which the velocity of
fluid in the interior turns to be negative, i.e.,
\begin{equation}\label{g15}
\mathbb{U}=\mathbb{D}_{\mathbb{T}}C< 0.
\end{equation}
Equations \eqref{g14} and \eqref{g15} provide the relationship
between the C-energy and velocity as
\begin{equation}\label{g16}
\mathbb{E}\equiv\frac{C'}{B}=\sqrt{\mathbb{U}^2+\frac{1}{4}+\frac{s^2}{C^2}-\frac{2\tilde{m}}{C}}.
\end{equation}
We use the definition of $\mathbb{D}_{\mathbb{T}}$ to express the
energy variation inside cylindrical object as
\begin{align}\nonumber
\mathbb{D}_{\mathbb{T}}\tilde{m}&=-4\pi\left[\left(\bar{P_{r}}+\mathcal{T}^{1(\mathcal{C})}_1-\frac{\bar{s}^2}{8\pi
C^4}+\mathcal{E}^{1(\mathcal{C})}_1\right)\mathbb{U}\right.\\\label{g17}
&+\left.\left(\bar{\varsigma}-\mathcal{T}^{1(\mathcal{C})}_0-\mathcal{E}^{1(\mathcal{C})}_0\right)\mathbb{E}\right]C^2
+\frac{\dot C}{8A}-\frac{s^2\dot{C}}{2AC^2},
\end{align}
while it comes out to be in terms of $\mathbb{D}_{\mathbb{R}}$ as
\begin{align}\nonumber
\mathbb{D}_{\mathbb{R}}\tilde{m}&=4\pi\left[\left(\bar{\mu}+\mathcal{T}^{0(\mathcal{C})}_{0}+\frac{\bar{s}^2}{8\pi
C^4}+\mathcal{E}^{0(\mathcal{C})}_0\right)+\frac{1}{32\pi
C^2}\right.\\\label{g18}
&+\left.\left(\bar{\varsigma}-\mathcal{T}^{1(\mathcal{C})}_{0}
-\mathcal{E}^{1(\mathcal{C})}_{0}\right)\frac{\mathbb{U}}{\mathbb{E}}\right]C^2+\frac{ss'}{CC'}-\frac{s^2}{2C^2},
\end{align}
which further yields
\begin{align}\nonumber
\frac{3\tilde{m}}{C^3}&=4\pi\left(\bar{\mu}+\mathcal{T}^{0(\mathcal{C})}_{0}+\frac{\bar{s}^2}{8\pi
C^4}+\mathcal{E}^{0(\mathcal{C})}_0\right)-\frac{4\pi}{C^3}\int^{r}_{0}
\left[\mathbb{D}_{\mathbb{R}}\left(\bar{\mu}+\mathcal{T}^{0(\mathcal{C})}_{0}+\frac{\bar{s}^2}{8\pi
C^4}\right.\right.\\\label{g19}
&+\left.\left.\mathcal{E}^{0(\mathcal{C})}_0\right)-\frac{3\mathbb{U}}{C\mathbb{E}}\left(\bar{\varsigma}-\mathcal{T}^{1(\mathcal{C})}_{0}
-\mathcal{E}^{1(\mathcal{C})}_{0}\right)\right]C'C^3dr+\frac{3}{8C^2}+\frac{3s^2}{2C^4}.
\end{align}
The term $\frac{3}{8C^2}$ in the above equation is obvious from
C-energy \eqref{g13}. The Weyl tensor
$(C^{\lambda}_{\varphi\vartheta\omega})$ gives the amount of stretch
by which a massive body educes nearby celestial structures due to
fluctuations in its gravitational field. There are two independent
components which completely define this tensor, namely magnetic and
electric parts which are expressed, respectively as
\begin{eqnarray}\nonumber
H_{\varphi\vartheta}&=&\frac{1}{2}\eta_{\varphi\omega\nu\gamma}C^{\nu\gamma}_{\vartheta\mu}\mathcal{K}^{\omega}
\mathcal{K}^{\mu},\\\nonumber
E_{\varphi\vartheta}&=&C_{\varphi\omega\vartheta\nu}\mathcal{K}^{\omega}\mathcal{K}^{\nu}.
\end{eqnarray}
For the general case of cylindrical fluid, the magnetic part is
non-vanishing and depends on a scalar, while it disappears in the
current (restricted) setup. Moreover, the electric part can also be
expressed in an alternative way as
\begin{equation}\label{g20}
E_{\varphi\vartheta}
=\varepsilon\bigg(\mathcal{W}_{\varphi}\mathcal{W}_{\vartheta}-\frac{h_{\varphi\vartheta}}{3}\bigg),
\end{equation}
where the value of electric scalar is
\begin{eqnarray}\nonumber
\varepsilon &=&\frac{1}{2A^2}\left[\frac{\ddot C}{C}-\frac{\ddot
B}{B}-\left(\frac{\dot C}{C}+\frac{\dot A}{A}\right)\left(\frac{\dot
C}{C}-\frac{\dot
B}{B}\right)\right]+\frac{1}{2B^2}\left[\left(\frac{C'}{C}
-\frac{A'}{A}\right)\right.\\\label{g21}
&\times&\left.\left(\frac{C'}{C}+\frac{B'}{B}\right)+\frac{A''}{A}
-\frac{C''}{C}\right]-\frac{1}{2C^2}.
\end{eqnarray}
We would like to point out here that we describe the electric Weyl
tensor in terms of a single scalar function due to the constrained
character of the cylinder, whereas for the general cylindrically
symmetric fluid, it is defined in the form of two scalars
\cite{36d}.

The effects of tidal force can also be described by the Weyl tensor,
and can be studied by the following equation due to the scalar
\eqref{g21} as
\begin{align}\nonumber
\frac{3\tilde{m}}{C^3}&=-\varepsilon+4\pi\left\{\left(\bar{\mu}+\mathcal{T}^{0(\mathcal{C})}_{0}+\frac{\bar{s}^2}{8\pi
C^4}+\mathcal{E}^{0(\mathcal{C})}_0\right)-\Pi^{(\mathrm{EFF})}-\mathcal{E}^{(\mathcal{C})}\right\}\\\label{g22}
&+\frac{3}{8C^2}+\frac{3s^2}{2C^4},
\end{align}
where
$\mathcal{E}^{(\mathcal{C})}=\mathcal{E}^{1(\mathcal{C})}_1-\mathcal{E}^{2(\mathcal{C})}_2,~\Pi^{(\mathrm{EFF})}=\bar{\Pi}+\Pi^{(\mathcal{C})},
~\bar{\Pi}=\frac{\Pi}{1-\frac{\Phi
s^2}{2C^4}}$ and
$\Pi^{(\mathcal{C})}=\mathcal{T}^{1(\mathcal{C})}_{1}-\mathcal{T}^{2(\mathcal{C})}_{2}$.

\section{Structure Scalars}

Herrera et al. \cite{41bb} proposed an extensive approach to split
the Riemann tensor orthogonally. We use this approach in
$f(\mathcal{R},\mathcal{T},\mathcal{R}_{\varphi\vartheta}\mathcal{T}^{\varphi\vartheta})$
gravity which yields some tensors and can further be split in their
trace and trace-free parts. These scalars must associate with
certain physical variables of the configuration. The Riemann tensor,
the Weyl tensor as well as modified $\mathbb{EMT}$ and its trace can
be interlinked through the following equation as
\begin{equation}\label{g23}
\mathcal{R}^{\omega\varphi}_{\gamma\vartheta}=C^{\omega\varphi}_{\gamma\vartheta}+16\pi
\Omega^{(\mathrm{EFF})[\omega}_{[\gamma}\delta^{\varphi]}_{\vartheta]}+8\pi
\Omega^{(\mathrm{EFF})}\left(\frac{1}{3}\delta^{\omega}_{[\gamma}\delta^{\varphi}_{\vartheta]}
-\delta^{[\omega}_{[\gamma}\delta^{\varphi]}_{\vartheta]}\right),
\end{equation}
where anti-symmetric property of the indices
$(\omega,\varphi,\gamma,\vartheta)$ is used. Here,
$\Omega^{(\mathrm{EFF})[\omega}_{[\gamma}\delta^{\varphi]}_{\vartheta]}=
\mathcal{T}^{(\mathrm{EFF})[\omega}_{[\gamma}\delta^{\varphi]}_{\vartheta]}+\mathcal{E}^{[\omega}_{[\gamma}\delta^{\varphi]}_{\vartheta]}$.
Equation \eqref{g23} produces certain tensors, i.e.,
$\mathcal{Y}_{\varphi\vartheta}$ and
$\mathcal{X}_{\varphi\vartheta}$ (after some lengthy calculations)
as
\begin{eqnarray}\label{g24}
\mathcal{Y}_{\varphi\vartheta}
&=&\mathcal{R}_{\varphi\omega\vartheta\gamma}\mathcal{K}^{\omega}\mathcal{K}^{\gamma},\\\label{g25}
\mathcal{X}_{\varphi\vartheta}&=&^{\ast}\mathcal{R}^{\ast}_{\varphi\omega\vartheta\gamma}\mathcal{K}^{\omega}\mathcal{K}^{\gamma}
=\frac{1}{2}\eta^{\epsilon\nu}_{\varphi\omega}\mathcal{R}^{\ast}_{\epsilon\nu\vartheta\gamma}\mathcal{K}^{\omega}\mathcal{K}^{\gamma},
\end{eqnarray}
where $\eta^{\epsilon\nu}_{\varphi\omega}$ and
$\mathcal{R}^{\ast}_{\varphi\omega\vartheta\gamma}$ are the
Levi-Civita symbol and the dual Riemann tensor, respectively,
defined as $\mathcal{R}^{\ast}_{\varphi\omega\vartheta\gamma}
=\frac{1}{2}\eta_{\omega\nu\vartheta\gamma}\mathcal{R}^{\omega\nu}_{\varphi\omega}$.
The alternate expressions of tensors \eqref{g24} and \eqref{g25} are
\begin{eqnarray}\label{g26}
\mathcal{Y}_{\varphi\vartheta}&=&\frac{h_{\varphi\vartheta}\mathcal{Y}_{T}}{3}+\bigg(\mathcal{W}_{\varphi}\mathcal{W}_{\vartheta}
-\frac{h_{\varphi\vartheta}}{3}\bigg)\mathcal{Y}_{TF},\\\label{g27}
\mathcal{X}_{\varphi\vartheta}&=&\frac{h_{\varphi\vartheta}\mathcal{X}_{T}}{3}+\bigg(\mathcal{W}_{\varphi}\mathcal{W}_{\vartheta}
-\frac{h_{\varphi\vartheta}}{3}\bigg)\mathcal{X}_{TF},
\end{eqnarray}
where $\mathcal{Y}_{T},~\mathcal{X}_{T},~\mathcal{Y}_{TF}$ and
$\mathcal{X}_{TF}$ are trace and trace-free parts, respectively. In
this scenario, these scalar functions come out to be
\begin{eqnarray}\label{g28}
&&\mathcal{X}_{T}=\frac{1}{1-\frac{\Phi
s^2}{2C^4}}\bigg(8\pi\mu-\frac{s^2}{C^4}\bigg)\bigg(\frac{1}{2}\Phi\mathcal{R}+1\bigg)
+\chi^{(\mathcal{C})}_{1},\\\label{g29}
&&\mathcal{X}_{TF}=-\varepsilon-\frac{1}{1-\frac{\Phi
s^2}{2C^4}}\bigg(4\pi\Pi-\frac{s^2}{C^4}\bigg)\bigg(\frac{1}{2}\Phi\mathcal{R}+1\bigg),\\\label{g30}
&&\mathcal{Y}_{T}=\frac{1}{1-\frac{\Phi
s^2}{2C^4}}\bigg\{4\pi(\mu+3P_{r}-2\Pi)+\frac{s^2}{C^4}\bigg\}\bigg(\frac{1}{2}\Phi\mathcal{R}+1\bigg)
+\chi^{(\mathcal{C})}_{2},\\\label{g31}
&&\mathcal{Y}_{TF}=\varepsilon-\frac{1}{1-\frac{\Phi
s^2}{2C^4}}\bigg(4\pi\Pi-\frac{s^2}{C^4}\bigg)\bigg(\frac{1}{2}\Phi\mathcal{R}+1\bigg)
+\chi^{(\mathcal{C})}_{3},
\end{eqnarray}
where the values of modified corrections $\chi^{(\mathcal{C})}_{1}$,
$\chi^{(\mathcal{\mathcal{C}})}_{2}$ and
$\chi^{(\mathcal{C})}_{3}=\frac{1}{\mathcal{W}_{\varphi}\mathcal{W}_{\vartheta}-\frac{1}{3}h_{\varphi\vartheta}}
\chi_{\varphi\vartheta}^{(\mathcal{C})}$ are given in Appendix
\textbf{B}. It has been mentioned earlier that we have imposed
restrictions on geometrical structure and the fluid distribution. As
a result, we obtain only one structure scalar corresponding to the
trace-free part of the electric component of the Riemann tensor
$\mathcal{Y}_{\varphi\vartheta}$. The scalar $\mathcal{X}_{T}$
incorporates only homogeneous energy density while factor
$\mathcal{Y}_{T}$ encompasses local anisotropic pressure as well
along with modified corrections.

The structural evolution of dynamical cylinder can be studied
through $\mathcal{Y}_{TF}$ (which guarantees involvement of the
inhomogeneous energy density and anisotropic pressure in the fluid
distribution) whose alternative form in terms of scalar \eqref{g21}
is
\begin{align}\nonumber
\mathcal{Y}_{TF}&=-8\pi\bar{\Pi}-4\pi\big(\Pi^{(\mathcal{C})}+\mathcal{E}^{(\mathcal{C})}\big)
+\chi^{(\mathcal{C})}_{3}-\frac{\Phi\mathcal{R}}{2}\bigg(4\pi\bar{\Pi}-\frac{\bar{s}^2}{C^4}\bigg)\\\nonumber
&+\frac{\bar{s}^2}{C^4}+\frac{4\pi}{C^3}\int_0^r
C^3\left[\mathbb{D}_{\mathbb{R}}\left(\bar{\mu}+\mathcal{T}^{0(\mathcal{C})}_{0}+\frac{\bar{s}^2}{8\pi
C^4}+\mathcal{E}^{0(\mathcal{C})}_{0}\right)\right.\\\label{g32}
&-\left.3\left(\bar{\varsigma}-\mathcal{T}^{1(\mathcal{C})}_{0}-\mathcal{E}^{1(\mathcal{C})}_{0}\right)\frac{\mathbb{U}}{C
\mathbb{E}}\right]C'dr.
\end{align}
It is observed from Eq.\eqref{g32} that modified scalar
$\mathcal{Y}_{TF}$ comprises all physical parameters such as
effective inhomogeneous energy density, dissipation flux, charge and
pressure anisotropy. Another factor appears in the orthogonal
splitting is $\mathcal{X}_{TF}$ \eqref{g29} that helps to analyze
the inhomogeneity of the energy density of fluid configuration as
\begin{align}\nonumber
\mathcal{X}_{TF}&=-\frac{4\pi}{C^3}\int_0^r
\left[\mathbb{D}_{\mathbb{R}}\left(\bar{\mu}+\mathcal{T}^{0(\mathcal{C})}_{0}+\frac{\bar{s}^2}{8\pi
C^4}+\mathcal{E}^{0(\mathcal{C})}_{0}\right)
-3\left(\bar{\varsigma}-\mathcal{T}^{1(\mathcal{C})}_{0}-\mathcal{E}^{1(\mathcal{C})}_{0}\right)\right.\\\label{g33}
&\times\left.\frac{\mathbb{U}}{C\mathbb{E}}\right]C^3C'dr+4\pi\big(\Pi^{(\mathcal{C})}+\mathcal{E}^{(\mathcal{C})}\big)
-\frac{\Phi\mathcal{R}}{2}\bigg(4\pi\bar{\Pi}-\frac{\bar{s}^2}{C^4}\bigg)+\frac{\bar{s}^2}{C^4}.
\end{align}

\section{Different Modes of Evolution}

The nature of any geometrical configuration (in the absence as well
as presence of charge) can be understood through the study of
several state variables such as energy density and radial/tangential
pressure. It is observed from Eq.\eqref{g32} that scalar
$\mathcal{Y}_{TF}$ entails the combination of all these quantities
together with dissipation flux and charge in association with
modified corrections. Thereby, we choose it as the complexity factor
for non-static cylindrical distribution influenced from
electromagnetic field. Subsequently, the condition
$\mathcal{Y}_{TF}=0$ leads to the complexity-free system. Two
evolutionary patterns (homologous evolution and homogeneous
expansion) are considered in the following subsections to examine
the dynamical changes in the interior of self-gravitating object. We
will construct some limitations which ultimately leads to the less
complex system throughout the evolution.

\subsection{Homologous Evolution}

The term homologous refers to the system which has same pattern
throughout. The core of a compact object becomes so heavy after the
inward fall of all the material into it, due to which that body
collapses. Nonetheless, the radial distance and velocity of the
fluid are directly related to each other in homologous collapse.
Thus we can say that the core attracts all the matter at the same
rate during the collapse which consequently emits much more
gravitational radiations dissimilar to the body whose core collapses
initially. Equation \eqref{g12} in terms of velocity of the fluid
has the form
\begin{equation}\label{g34}
\mathbb{D}_{\mathbb{R}}\bigg(\frac{\mathbb{U}}{C}\bigg)=\frac{1}{C}\bigg\{\sigma+\frac{4\pi
C}{\mathbb{E}}\bigg(\bar{\varsigma}-\mathcal{T}^{1(\mathcal{C})}_{0}-\mathcal{E}^{1(\mathcal{C})}_{0}\bigg)\bigg\}.
\end{equation}
After integrating Eq.\eqref{g34}, we have
\begin{equation}\label{g35}
\mathbb{U}=\mathrm{x}(t)C+C\int^{r}_{0}\frac{C'}{C}\bigg\{\sigma+\frac{4\pi
C}{\mathbb{E}}\bigg(\bar{\varsigma}-\mathcal{T}^{1(\mathcal{C})}_{0}-\mathcal{E}^{1(\mathcal{C})}_{0}\bigg)\bigg\}dr,
\end{equation}
where $\mathrm{x}(t)$ serves as an integration function. The final
form of the velocity of collapsing cylindrical distribution at the
boundary can be determined as
\begin{equation}\label{g36}
\mathbb{U}=C\left[\frac{\mathbb{U}_{\Sigma}}{C_{\Sigma}}-\int^{r_{\Sigma}}_{r}
\frac{C'}{C}\bigg\{\sigma+\frac{4\pi
C}{\mathbb{E}}\bigg(\bar{\varsigma}-\mathcal{T}^{1(\mathcal{C})}_{0}-\mathcal{E}^{1(\mathcal{C})}_{0}\bigg)\bigg\}dr\right].
\end{equation}
The deviation of cylindrical structure from homologous mode can be
studied through some significant factors, i.e., heat dissipation and
shear scalar. We can observe homologous evolution inside the system
\cite{42bc,42bd}, if effects of the above integrand disappears.
Consequently, Eq.\eqref{g36} is left with $\mathbb{U}\sim C$ leads
to $\mathbb{U}=\mathrm{x}(t)C$ and
$\mathrm{x}(t)=\frac{\mathbb{U}_{\Sigma}}{C_{\Sigma}}$. The
homologous condition for the fluid influenced from electromagnetic
field has the form
\begin{equation}\label{g37}
\frac{4\pi
BC}{C'}\big(\bar{\varsigma}-\mathcal{T}^{1(\mathcal{C})}_{0}-\mathcal{E}^{1(\mathcal{C})}_{0}\big)+\sigma=0.
\end{equation}

\subsection{Homogeneous Expansion}

The constraint $\Theta'=0$ is required to discuss another
phenomenon, called as homogeneous expansion. This phase takes place
when the rate at which cosmic bodies collapse or expand is not
dependent on $r$, unlike preceding mode. This constraint becomes
together with Eq.\eqref{g12} as
\begin{equation}\label{g38}
4\pi\big(\bar{\varsigma}-\mathcal{T}^{1(\mathcal{C})}_{0}-\mathcal{E}^{1(\mathcal{C})}_{0}\big)=-\frac{C'}{3B}
\bigg\{\frac{3\sigma}{C}+\mathbb{D}_{\mathbb{R}}(\sigma)\bigg\}.
\end{equation}
The simultaneous use of the homologous condition \eqref{g37} and
Eq.\eqref{g38} yields $\mathbb{D}_{\mathbb{R}}(\sigma)=0$. Due to
the regularity condition at the core, we have $\sigma=0$ which makes
Eq.\eqref{g38} as
\begin{equation}\label{g39}
\bar{\varsigma}=\mathcal{T}^{1(\mathcal{C})}_{0}+\mathcal{E}^{1(\mathcal{C})}_{0},
\end{equation}
which discloses the incorporation of dissipative effects due to
$f(\mathcal{R},\mathcal{T},\mathcal{R}_{\varphi\vartheta}\mathcal{T}^{\varphi\vartheta})$
corrections and thus opposing $\mathbb{GR}$, where the homogeneous
evolution results in non-dissipative and shear-free matter source
\cite{33}.

\section{Some Kinematical and Dynamical Considerations}

In this section, some physical entities are analyzed to choose the
simplest possible evolutionary mode. Equation \eqref{g12} along with
the homologous condition \eqref{g37} give
\begin{equation}\label{g40}
\left(\Theta-\sigma\right)'=\left(\frac{3\dot C}{AC}\right)'=0.
\end{equation}
We take the metric potential $C(t,r)$ as a separable function of
both coordinates. Thus we obtain $A'=0$ \big(i.e., $a=0$ from
Eq.\eqref{g11e}\big) leads to the geodesic fluid. Further, we put
$A=1$ without any loss of generality. On the contrary,
Eqs.\eqref{g11b} and \eqref{g11d} for $A=1$ yields
\begin{equation}\label{g41}
\Theta-\sigma=\frac{3\dot C}{C},
\end{equation}
which provides $(\Theta-\sigma)'=0$, and hence recovering the
homologous condition \eqref{g40}. Thus the necessary and sufficient
condition for the dynamical cylinder to evolve in homologous mode is
that the fluid must follow geodesic path. In $\mathbb{GR}$, the
absence of heat dissipation
($\varsigma=0\Rightarrow\bar{\varsigma}=0$) implies the
disappearance of the shear ($\sigma=0$) in the matter source as
opposed to $f(\mathcal{R},\mathcal{T},\mathcal{Q})$ framework where
we have
\begin{equation}\label{g42}
\sigma=\frac{4\pi
BC}{C'}\big(\mathcal{T}^{1(\mathcal{C})}_{0}+\mathcal{E}^{1(\mathcal{C})}_{0}\big).
\end{equation}
Equation \eqref{g38} produces the shear scalar corresponding to
homogeneous pattern as
\begin{equation}\label{g43}
\sigma=\frac{\mathrm{y}(t)}{C^3}+\frac{12\pi}{C^3}\int_{0}^{r}
BC^3\big(\mathcal{T}^{1(\mathcal{C})}_{0}+\mathcal{E}^{1(\mathcal{C})}_{0}\big)dr,
\end{equation}
where $\mathrm{y}(t)$ is an arbitrary integration function.
Furthermore, it is deduced that homogenous pattern implies
homologous condition ($\sigma=0 \Rightarrow U \sim C$) when heat
dissipation and modified corrections are neglected. For the current
cylindrical setup, the C-energy can be linked with collapse rate as
\begin{equation}\label{g44}
\mathbb{D}_{\mathbb{T}}\mathbb{U}=-\frac{\tilde{m}}{C^2}-4\pi
C\left(\bar{P_{r}}+\mathcal{T}^{1(\mathcal{C})}_{1}-\frac{\bar{s}^2}{8\pi
C^4}+\mathcal{E}^{1(\mathcal{C})}_{1}\right)+\frac{1}{8C}+\frac{s^2}{2C^3}.
\end{equation}
We combine Eq.\eqref{g44} with the scalar $\mathcal{Y}_{TF}$
\eqref{g31} to get
\begin{align}\nonumber
\frac{3\mathbb{D}_{\mathbb{T}}\mathbb{U}}{C}&=\mathcal{Y}_{TF}+\frac{\Phi\mathcal{R}}{2}\bigg(4\pi\bar{\Pi}-\frac{\bar{s}^2}{C^4}\bigg)
-4\pi\big\{\bar{\mu}+\mathcal{T}^{0(\mathcal{C})}_{0}+\mathcal{E}^{0(\mathcal{C})}_{0}\\\label{g45}
&+3\big(\bar{P_{r}}+\mathcal{T}^{1(\mathcal{C})}_{1}+\mathcal{E}^{1(\mathcal{C})}_{1}\big)-2\bar{\Pi}-\Pi^{(\mathcal{C})}
-\mathcal{E}^{(\mathcal{C})}\big\}-\chi_3^{(\mathcal{C})}.
\end{align}
Using Eqs.\eqref{g8}, \eqref{g8b} and \eqref{g8c}, we have
\begin{align}\nonumber
&4\pi\big\{\bar{\mu}+\mathcal{T}^{0(\mathcal{C})}_{0}+\mathcal{E}^{0(\mathcal{C})}_{0}
-2\big(\bar{\Pi}-\Pi^{(\mathcal{C})}-\mathcal{E}^{(\mathcal{C})}\big)\\\label{g46}
&+3\big(\bar{P_{r}}+\mathcal{T}^{1(\mathcal{C})}_{1}+\mathcal{E}^{1(\mathcal{C})}_{1}\big)\big\}=-\frac{2\ddot
C}{C}-\frac{\ddot B}{B}-\frac{\bar{s}^2}{C^4},
\end{align}
and
\begin{equation}\label{g47}
\frac{3\mathbb{D}_{\mathbb{T}}\mathbb{U}}{C}=\frac{3\ddot C}{C}.
\end{equation}
Equations \eqref{g45}-\eqref{g47} simultaneously lead to
\begin{equation}\label{g48}
\frac{\ddot C}{C}-\frac{\ddot
B}{B}=\mathcal{Y}_{TF}-\chi_3^{(\mathcal{C})}-4\pi\Pi^{(\mathcal{C})}-4\pi\mathcal{E}^{(\mathcal{C})}+\frac{\bar{s}^2}{C^4}
+\frac{\Phi\mathcal{R}}{2}\bigg(4\pi\bar{\Pi}-\frac{\bar{s}^2}{C^4}\bigg).
\end{equation}
It should be mentioned here that the complexity-free structure can
be obtained by taking $\frac{\ddot C}{C}-\frac{\ddot
B}{B}+\chi_3^{(\mathcal{C})}+4\pi\Pi^{(\mathcal{C})}+4\pi\mathcal{E}^{(\mathcal{C})}-\frac{\bar{s}^2}{C^4}
-\frac{\Phi\mathcal{R}}{2}\big(4\pi\bar{\Pi}-\frac{\bar{s}^2}{C^4}\big)=0$.

The metric potentials $A,~B$ and $C$ are unknown quantities that are
needed to calculate. As we have assumed $A=1$, thus only $B$ and $C$
are left whose possible solutions can be obtained by solving two
equations. In this scenario, we consider complexity-free
($\mathcal{Y}_{TF}=0$) and homologous conditions to evaluate these
unknowns with and without considering the effects of heat
dissipation. These conditions, in the absence of dissipative flux
are presented as Eqs.\eqref{g51} and \eqref{g52}. The
complexity-free condition remains the same in the presence of
dissipation whereas Eq.\eqref{g53} provides the homologous
condition.

\section{Stability of Complexity-free Condition}

It is possible that an interior configuration is complexity-free at
some initial time and develops complex nature at a later time during
the evolution due to the involvement of some factors. To investigate
these quantities, we extract the evolution equation for scalar
$\mathcal{Y}_{TF}$ with the help of a standard technique \cite{41bb}
along with Eqs.\eqref{g11}, \eqref{g29} and \eqref{g31} as
\begin{align}\nonumber
&-\frac{4\pi}{B}\bigg\{\bar{\varsigma}'-\frac{C'}{C}\left(\bar{\varsigma}+3\mathcal{T}^{1(\mathcal{C})}_{0}
+3\mathcal{E}^{1(\mathcal{C})}_{0}\right)\bigg\}-4\pi\left(\bar{\mu}+\bar{p_{r}}\right)\sigma-\dot{\mathcal{Y}}_{TF}+\dot\chi_3^{(\mathcal{C})}
-8\pi\dot{\bar{\Pi}}\\\nonumber
&-4\pi\dot{\Pi}^{(\mathcal{C})}-4\pi\mathbb{Z}_1+4\pi\dot{\mathcal{T}}^{0(\mathcal{C})}_{0}+4\pi\dot{\mathcal{E}}^{0(\mathcal{C})}_{0}
-4\pi\dot{\mathcal{E}}^{(\mathcal{C})}-\frac{\Phi}{2}\bigg\{\mathcal{R}\bigg(4\pi\bar{\Pi}-\frac{\bar{s}^2}{C^4}\bigg)\bigg\}^.\\\nonumber
&+\frac{\Phi}{16\pi+\Phi\mathcal{R}}\big(\mu\dot{\mathcal{R}}+\frac{\varsigma\mathcal{R}'}{B}\big)-\frac{3\dot{C}}{C}
\bigg\{\mathcal{Y}_{TF}-\frac{\bar{s}^2}{C^4}+\frac{\Phi\mathcal{R}}{2}\bigg(4\pi\bar{\Pi}-\frac{\bar{s}^2}{C^4}\bigg)
-\chi_3^{(\mathcal{C})}\bigg\}\\\label{g54}
&+\frac{12\pi\dot{C}}{C}\left(\mathcal{T}^{0(\mathcal{C})}_{0}+\mathcal{T}^{2(\mathcal{C})}_{2}
+\mathcal{E}^{0(\mathcal{C})}_{0}+\mathcal{E}^{2(\mathcal{C})}_{2}\right)-\frac{16\pi\bar{\Pi}\dot{C}}{C}=0.
\end{align}
Firstly, for the non-dissipative case, we take
$\mathcal{Y}_{TF}=0=\Pi^{(\mathrm{EFF})}=\varsigma=\sigma$ at some
initial time (say, $t=0$) for which Eq.\eqref{g54} yields
\begin{align}\nonumber
&\frac{12\pi{C'}}{BC}\left(\mathcal{T}^{1(\mathcal{C})}_{0}+\mathcal{E}^{1(\mathcal{C})}_{0}\right)
+\frac{\Phi}{2}\bigg(\frac{\mathcal{R}\bar{s}^2}{C^4}\bigg)^.+\frac{\Phi\mu\dot{\mathcal{R}}}{16\pi+\Phi\mathcal{R}}
-\dot{\mathcal{Y}}_{TF}+\dot\chi_3^{(\mathcal{C})}+4\pi\dot{\mathcal{E}}^{0(\mathcal{C})}_{0}\\\nonumber
&+\frac{3\dot{C}}{C}\bigg(\frac{\bar{s}^2}{C^4}+\frac{\Phi\mathcal{R}\bar{s}^2}{C^4}+\chi_3^{(\mathcal{C})}\bigg)
-4\pi\dot{\mathcal{E}}^{(\mathcal{C})}-8\pi\dot{\bar{\Pi}}-4\pi\dot{\Pi}^{(\mathcal{C})}-4\pi\mathbb{Z}_1
+4\pi\dot{\mathcal{T}}^{0(\mathcal{C})}_{0}\\\label{g55}
&+\frac{12\pi\dot{C}}{C}\left(\mathcal{T}^{0(\mathcal{C})}_{0}+\mathcal{T}^{2(\mathcal{C})}_{2}
+\mathcal{E}^{0(\mathcal{C})}_{0}+\mathcal{E}^{2(\mathcal{C})}_{2}\right)=0.
\end{align}
Together use of this equation and the time derivative of
Eq.\eqref{g32} (at $t=0$), we have
\begin{align}\nonumber
&4\pi\frac{\partial}{\partial
t}\bigg[\frac{1}{C^3}\int^{r}_{0}C^3\bigg\{\left(\bar{\mu}+\mathcal{T}^{0(\mathcal{C})}_{0}+\frac{\bar{s}^2}{C^4}
+\mathcal{E}^{0(\mathcal{C})}_{0}\right)'+\frac{3B\dot{C}}{C}\left(\mathcal{T}^{1(\mathcal{C})}_{0}
+\mathcal{E}^{1(\mathcal{C})}_{0}\right)\bigg\}dr\bigg]\\\nonumber
&=4\pi\dot{\mathcal{T}}^{0(\mathcal{C})}_{0}+2\pi\Phi\big(\bar{\Pi}\mathcal{R}\big)^.-4\pi\dot{\mathcal{E}}^{0(\mathcal{C})}_{0}
+8\pi\dot{\mathcal{E}}^{(\mathcal{C})}+\frac{4\bar{s}^2\dot{C}}{C^5}
+\frac{12\pi{C'}}{BC}\left(\mathcal{T}^{1(\mathcal{C})}_{0}+\mathcal{E}^{1(\mathcal{C})}_{0}\right)\\\nonumber
&+\frac{3\dot{C}}{C}\bigg(\frac{\bar{s}^2}{C^4}+\frac{\Phi\mathcal{R}\bar{s}^2}{C^4}+\chi_3^{(\mathcal{C})}\bigg)
+\frac{12\pi\dot{C}}{C}\left(\mathcal{T}^{0(\mathcal{C})}_{0}+\mathcal{T}^{2(\mathcal{C})}_{2}
+\mathcal{E}^{0(\mathcal{C})}_{0}+\mathcal{E}^{2(\mathcal{C})}_{2}\right)\\\label{g56}
&-4\pi\mathbb{Z}_1+\frac{\Phi\mu\dot{\mathcal{R}}}{16\pi+\Phi\mathcal{R}}.
\end{align}
The stability of $\mathcal{Y}_{TF}=0$ depends on state determinants
(density and pressure). We see from Eq.\eqref{g56} that the system
could depart from stability if the interior configuration involves
inhomogeneous energy density, pressure anisotropy and charge. Thus,
in this scenario of charged configuration, the electromagnetic field
also disturbs stability of the considered setup. Moreover, by
substituting $\sigma=\mathcal{Y}_{TF}=0$ in Eq.\eqref{g54}, we have
the most general case involving dissipation flux as
\begin{align}\nonumber
&-\frac{4\pi}{B}\bigg\{\bar{\varsigma}'-\frac{C'}{C}\left(\bar{\varsigma}+3\mathcal{T}^{1(\mathcal{C})}_{0}
+3\mathcal{E}^{1(\mathcal{C})}_{0}\right)\bigg\}+\frac{\Phi}{16\pi+\Phi\mathcal{R}}\left(\mu\dot{\mathcal{R}}
+\frac{\varsigma\mathcal{R}'}{B}\right)-\dot{\mathcal{Y}}_{TF}\\\nonumber
&+\frac{\Phi}{2}\bigg(\frac{\mathcal{R}\bar{s}^2}{C^4}\bigg)^.+\frac{3\dot{C}}{C}\bigg(\frac{\bar{s}^2}{C^4}+\frac{\Phi\mathcal{R}\bar{s}^2}{C^4}
+\chi_3^{(\mathcal{C})}\bigg)+\dot\chi_3^{(\mathcal{C})}+4\pi\dot{\mathcal{E}}^{0(\mathcal{C})}_{0}
-4\pi\dot{\mathcal{E}}^{(\mathcal{C})}-8\pi\dot{\bar{\Pi}}\\\label{g57}
&-4\pi\dot{\Pi}^{(\mathcal{C})}-4\pi\mathbb{Z}_1+4\pi\dot{\mathcal{T}}^{0(\mathcal{C})}_{0}
+\frac{12\pi\dot{C}}{C}\left(\mathcal{T}^{0(\mathcal{C})}_{0}+\mathcal{T}^{2(\mathcal{C})}_{2}
+\mathcal{E}^{0(\mathcal{C})}_{0}+\mathcal{E}^{2(\mathcal{C})}_{2}\right)=0.
\end{align}
This shows that stability is now affected also by the heat flux.

\section{Final Remarks}

Our universe contains plenty of astronomical bodies whose
astonishing nature prompted many researchers to study their complex
structures. This paper is devoted to studying different physical
variables representing interior of the cylindrical fluid
distribution that make the celestial object more complex in
$f(\mathcal{R},\mathcal{T},\mathcal{R}_{\varphi\vartheta}\mathcal{T}^{\varphi\vartheta})$
framework. A standard model $\mathcal{R}+\Phi
\mathcal{R}_{\varphi\vartheta}\mathcal{T}^{\varphi\vartheta}$ has
been considered in this regard to study the matter-geometry coupling
effects on non-static spacetime influenced from electromagnetic
field. We have considered anisotropic matter distribution coupled
with heat flux. The Riemann tensor has been split orthogonally
through Herrera's technique which resulted in four scalars, each of
them is uniquely defined in terms of particular physical parameters.
Following reasons justify the adoption of $\mathcal{Y}_{TF}$ from
four resulting candidates as the complexity factor.
\begin{enumerate}
\item This factor has been recognized as the best candidate for the complexity factor in $\mathbb{GR}$ \cite{31} as well
as
$f(\mathcal{R},\mathcal{T},\mathcal{R}_{\varphi\vartheta}\mathcal{T}^{\varphi\vartheta})$
gravity \cite{26d,26e} for static uncharged/charged spacetimes.
Thus, $\mathcal{Y}_{TF}$ can be recovered from Eq.\eqref{g31} for
static scenario.
\item All physical quantities such as energy density
inhomogeneity, heat dissipation, pressure anisotropy and charge
together with correction terms should be included in complexity
factor which is ensured only by this factor.
\end{enumerate}
We have considered two simplest evolutionary patterns for
self-gravitating dynamical structure, i.e., homogeneous expansion
and homologous evolution, and studied them in modified gravity. We
have constructed the possible solutions for metric potentials in the
case of dissipation as well as non-dissipation with the help of
homologous condition \eqref{g37} and $\mathcal{Y}_{TF}=0$. We have
also discussed several factors under which the system deviates from
complexity-free scenario.

It has been noticed that the strong non-minimally coupled model
\eqref{g5d} makes dynamical cylinder more complex due to
incorporation of product terms of the matter variables and metric
potentials. We have considered the homologous fluid to be of
geodesic nature (i.e., $A=1$), and thus this mode was suggested as
the simplest pattern of evolution. The fulfilment of the requirement
$\mu=\varsigma=\Pi=0$ provides complexity-free structure
($\mathcal{Y}_{TF}=0$) in $\mathbb{GR}$, while an additional
condition
($8\pi\bar{\Pi}+4\pi\Pi^{(\mathcal{C})}+2\pi\zeta\mathcal{R}\bar{\Pi}-\chi_3^{(\mathcal{C})}=0$)
is needed to disappear the complexity factor in this gravity. It is
found that this modified theory does not provide shear-free
($\sigma=0$) structure even in non-dissipative case ($\varsigma=0$),
in contrast to $\mathbb{GR}$, thus the dynamical cylinder becomes
more complex in the presence of charge and modified corrections.
This phenomenon has been investigated for minimal/non-minimal
$f(\mathcal{R},\mathcal{T})$ models. The compatibility of the
simplest evolutionary modes with each other has been observed for
the case of minimal coupling, and otherwise, not \cite{36a}. The
results we obtained for the considered
$f(\mathcal{R},\mathcal{T},\mathcal{Q})$ model are compatible with
those of $f(\mathcal{R},\mathcal{T})$ gravity. We have analyzed the
stability of vanishing complexity factor and figured out some
factors that enforced the system to deviate from its stability. It
is worth mentioning here that all our results can be recovered in
$\mathbb{GR}$ \cite{33} by vanishing the coupling parameter $\Phi$
and charge.

\vspace{0.25cm}

\section*{Appendix A}

\renewcommand{\theequation}{A\arabic{equation}}
\setcounter{equation}{0} The
$f(\mathcal{R},\mathcal{T},\mathcal{Q})$ corrections in the field
equations \eqref{g8}-\eqref{g8c} are
\begin{align}\nonumber
\mathcal{T}^{(\mathcal{C})}_{00}&=\frac{\Phi}{8\pi\big(1-\frac{\Phi
s^2}{2C^4}\big)}\bigg\{\mu\bigg(\frac{3\ddot{B}}{2B^2}-\frac{3\dot{A}\dot{B}}{2AB}
+\frac{3\ddot{C}}{C}-\frac{\dot{A}\dot{C}}{AC}-\frac{\dot{C}^2}{C^2}-\frac{3AA''}{2B^2}+\frac{2A'^2}{B^2}\\\nonumber
&-\frac{1}{2}A^2\mathcal{R}-\frac{AA'B'}{2B^3}-\frac{2\dot{B}\dot{C}}{BC}-\frac{3AA'C'}{B^2C}\bigg)-\dot{\mu}\bigg(\frac{3\dot{A}}{A}
+\frac{\dot{C}}{C}+\frac{\dot{B}}{2B}\bigg)+\frac{\mu''A^2}{2B^2}\\\nonumber
&-\mu'\bigg(\frac{A^2B'}{2B^3}-\frac{A^2C'}{B^2}\bigg)+P_{r}\bigg(-\frac{\ddot{B}}{2B}+\frac{AA''}{2B^2}-\frac{AA'B'}{2B^3}-\frac{A^2C''}{B^2C}
-\frac{A^2C'^2}{B^2C^2}\\\nonumber
&+\frac{2A^2B'C'}{B^3C}-\frac{\dot{B}\dot{C}}{BC}-\frac{2A^2B'\dot{C}}{B^3C}+\frac{2A'^2}{B^2}\bigg)+\frac{\dot{P}_{r}\dot{B}}{2B}
+P'_{r}\bigg(\frac{A^2B'}{2B^3}-\frac{2A^2C'}{B^2C}\bigg)\\\nonumber
&-\frac{P''_{r}A}{2B^2}-P_{\bot}\bigg(\frac{\ddot{C}}{C}+\frac{\dot{C}^2}{C}-\frac{\dot{A}\dot{C}}{AC}
-\frac{AA'C'}{B^2C}-\frac{A^2C''}{B^2C}+\frac{A^2B'C'}{B^3C}-\frac{A^2C'^2}{B^2C^2}\\\nonumber
&+\frac{\dot{B}\dot{C}}{BC}\bigg)-\frac{3\dot{P}_{\bot}\dot{C}}{C}+\frac{P'_{\bot}A^2C'}{B^2C}-\varsigma\bigg(\frac{9\dot{A}A'}{2AB}
-\frac{2A\dot{C}'}{BC}+\frac{3A'\dot{C}}{BC}+\frac{5A\dot{B}C'}{B^2C}\\\label{A1}
&+\frac{3A'\dot{B}}{B^2}+\frac{\dot{A}C'}{BC}+\frac{2A\dot{C}C'}{BC^2}\bigg)+\frac{2\dot{\varsigma}A'}{B}-\frac{2\varsigma'A\dot{C}}{BC}
+\frac{A^2\mathcal{Q}}{2}\bigg\},\\\nonumber
\mathcal{T}^{(\mathcal{C})}_{01}&=\frac{\Phi}{8\pi\big(1-\frac{\Phi
s^2}{2C^4}\big)}\bigg\{\mu\bigg(-\frac{\dot{A}A'}{A^2}-\frac{3\dot{A}'}{2A}
+\frac{2\dot{C}'}{C}-\frac{2\dot{B}C'}{BC}-\frac{2A'\dot{C}}{AC}\bigg)-\frac{3\dot{\mu}A'}{2A}\\\nonumber
&+\mu'\bigg(\frac{\dot{C}}{C}-\frac{\dot{B}}{2B}\bigg)+\frac{\dot{\mu}'}{2}+P_{r}\bigg(\frac{A'^2}{2B^2}-\frac{A'\dot{B}}{2AB}
-\frac{2\dot{C}'}{C}+\frac{2\dot{B}C'}{BC}+\frac{2A'\dot{C}}{AC}\bigg)\\\nonumber
&+\dot{P}_{r}\bigg(\frac{A'}{2A}-\frac{C'}{C}\bigg)+\frac{P'_{r}\dot{B}}{2B}-\frac{\dot{P}'_{r}}{2}+\frac{\dot{P}_{\bot}C'}{C}
+\frac{P'_{\bot}\dot{C}}{C}+\varsigma\bigg(\frac{1}{2}AB\mathcal{R}-\frac{2B\ddot{C}}{AC}\\\label{A2}
&-\frac{\ddot{B}}{A}+\frac{A''}{B}+\frac{2B\dot{A}\dot{C}}{A^2C}+\frac{2AC''}{BC}-\frac{2AB'C'}{B^2C}+\frac{\dot{A}\dot{B}}{A^2}
-\frac{A'B'}{B^2}\bigg)\bigg\},\\\nonumber
\mathcal{T}^{(\mathcal{C})}_{11}&=\frac{\Phi}{8\pi\big(1-\frac{\Phi
s^2}{2C^4}\big)}\bigg\{\mu\bigg(\frac{B\ddot{B}}{2A^2}-\frac{B\dot{A}\dot{B}}{2A^3}
-\frac{B^2\ddot{C}}{A^2C}+\frac{B^2\dot{A}\dot{C}}{A^3C}-\frac{B^2\dot{C}^2}{A^2C^2}-\frac{A''}{2A}\\\nonumber
&+\frac{A'B'}{2AB}-\frac{A'C'}{AC}\bigg)+\dot{\mu}\bigg(\frac{B^2\dot{A}}{2A^3}-\frac{2B^2\dot{C}}{A^2C}\bigg)
+\frac{\mu'A'}{2A}-\frac{\ddot{\mu}B^2}{2A^2}+P_{r}\bigg(\frac{1}{2}B^2\mathcal{R}\\\nonumber
&+\frac{3B\dot{A}\dot{B}}{2A^3}-\frac{3B\ddot{B}}{2A^2}-\frac{2A'C'}{AC}+\frac{3A''}{2A}-\frac{3A'B'}{2AB}-\frac{2B'C'}{BC}
+\frac{3C''}{C}-\frac{3B\dot{B}\dot{C}}{A^2C}\\\nonumber
&-\frac{2B'\dot{C}}{BC}-\frac{C'^2}{C^2}\bigg)+\dot{P}_{r}\bigg(\frac{B^2\dot{C}}{A^2C}-\frac{B^2\dot{A}}{2A^3}\bigg)
-P'_{r}\bigg(\frac{A'}{2A}+\frac{C'}{C}\bigg)+\frac{\ddot{P}_{r}B^2}{2A^2}\\\nonumber
&+P_{\bot}\bigg(\frac{A'C'}{AC}-\frac{B^2\ddot{C}}{A^2C}-\frac{B^2\dot{C}^2}{A^2C^2}+\frac{B^2\dot{A}\dot{C}}{A^3C}+\frac{C''}{C}
-\frac{B'C'}{BC}+\frac{C'^2}{C^2}-\frac{B\dot{B}\dot{C}}{A^2C}\bigg)\\\nonumber
&-\frac{\dot{P}_{\bot}B^2\dot{C}}{A^2C}-\frac{P'_{\bot}C'}{C}+\varsigma\bigg(\frac{2B\dot{C}'}{AC}-\frac{B\dot{A}C'}{A^2C}
-\frac{3BA'\dot{C}}{A^2C}-\frac{2B\dot{C}C'}{AC^2}-\frac{4\dot{B}C'}{AC}\bigg)\\\label{A3}
&+\frac{B^2\mathcal{Q}}{2}-\frac{2\dot{\varsigma}BC'}{AC}\bigg\},\\\nonumber
\mathcal{T}^{(\mathcal{C})}_{22}&=\frac{\Phi}{8\pi\big(1-\frac{\Phi
s^2}{2C^4}\big)}\bigg\{\mu\bigg(-\frac{C^2\ddot{B}}{2A^2B}+\frac{C^2\dot{A}\dot{B}}{2A^3B}
-\frac{C^2A''}{2AB^2}+\frac{C^2A'B'}{2AB^3}-\frac{C\dot{B}\dot{C}}{A^2B}\\\nonumber
&-\frac{CA'C'}{AB^2}\bigg)+\dot{\mu}\bigg(\frac{C^2\dot{A}}{2A^3}-\frac{C^2\dot{B}}{A^2B}-\frac{C\dot{C}}{A^2}\bigg)
-\frac{\mu'C^2A'}{2AB^2}-\frac{\ddot{\mu}C^2}{2A^2}+P_{r}\bigg(\frac{C^2\dot{A}\dot{B}}{2A^3B}\\\nonumber
&-\frac{C^2\ddot{B}}{2A^2B}-\frac{CA'C'}{AB^2}-\frac{C^2A''}{2AB^2}+\frac{C^2A'B'}{2AB^3}+\frac{CB'C'}{B^3}-\frac{C\dot{B}\dot{C}}{A^2B}
-\frac{2CB'\dot{C}}{B^3}\bigg)\\\nonumber
&-\frac{\dot{P}_{r}C^2\dot{B}}{2A^2B}+P'_{r}\bigg(-\frac{C^2A'}{AB^2}+\frac{C^2B'}{2B^3}-\frac{CC'}{B^2}\bigg)-\frac{P''_{r}C^2}{2B^2}
+P_{\bot}\bigg(\frac{1}{2}C^2\mathcal{R}\\\nonumber
&+\frac{2CC''}{B^2}-\frac{2C\ddot{C}}{A^2}-\frac{2\dot{C}^2}{A^2}+\frac{2C'^2}{B^2}-\frac{2CB'C'}{B^3}+\frac{2CA'C'}{AB^2}
-\frac{2C\dot{B}\dot{C}}{A^2B}-2\\\nonumber
&+\frac{2C\dot{A}\dot{C}}{A^3}\bigg)+\dot{P}_{\bot}C^2\bigg(\frac{\dot{B}}{2A^2B}-\frac{\dot{A}}{2A^3}\bigg)+P'_{\bot}C^2\bigg(-\frac{A'}{2AB^2}
+\frac{B'}{2B^3}\bigg)\\\nonumber
&+\frac{\ddot{P}_{\bot}C^2}{2A^2}-\frac{P''_{\bot}C^2}{2B^2}+\varsigma\bigg(-\frac{C^2\dot{A}'}{A^2B}+\frac{C^2\dot{A}A'}{A^3B}
-\frac{C\dot{A}C'}{A^2B}+\frac{C^2\dot{B}B'}{AB^3}-\frac{C^2\dot{B}'}{AB^2}\\\nonumber
&-\frac{CA'\dot{C}}{A^2B}-\frac{2C\dot{B}C'}{AB^2}\bigg)-\dot{\varsigma}\bigg(\frac{C^2A'}{A^2B}+\frac{CC'}{AB}\bigg)
-\varsigma'\bigg(\frac{C^2\dot{B}}{AB^2}+\frac{C\dot{C}}{AB}\bigg)-\frac{\dot{\varsigma}'C^2}{AB}\\\label{A4}
&+\frac{C^2\mathcal{Q}}{2}\bigg\},\\\nonumber
\mathcal{E}^{(\mathcal{C})}_{00}&=\frac{\Phi s^2}{8\pi
AB^3C^5\big(1-\frac{\Phi
s^2}{2C^4}\big)}\bigg\{2A^3BC''+A^2BCA''-2AB^2\dot{B}\dot{C}-2A^3B'C'\\\label{A5}
&-A^2CA'B'-AB^2C\ddot{B}+B^2C\dot{A}\dot{B}+\frac{A^3B^3C\mathcal{R}}{2}\bigg\},\\\label{A6}
\mathcal{E}^{(\mathcal{C})}_{01}&=\frac{\Phi s^2}{8\pi
ABC^5\big(1-\frac{\Phi
s^2}{2C^4}\big)}\big\{AB\dot{C}'-BA'\dot{C}-A\dot{B}C'\big\},\\\nonumber
\mathcal{E}^{(\mathcal{C})}_{11}&=\frac{\Phi s^2}{8\pi
A^3BC^5\big(1-\frac{\Phi
s^2}{2C^4}\big)}\bigg\{2AB^3\ddot{C}+AB^2C\ddot{B}-A^2BCA''-2A^2BA'C'\\\label{A7}
&-2B^3\dot{A}\dot{C}+A^2CA'B'-B^2C\dot{A}\dot{B}-\frac{A^3B^3C\mathcal{R}}{2}\bigg\},\\\label{A8}
\mathcal{E}^{(\mathcal{C})}_{22}&=\frac{\Phi s^2\mathcal{R}}{16\pi
C^2\big(1-\frac{\Phi s^2}{2C^4}\big)}.
\end{align}

\noindent The terms $\mathbb{Z}_1$ and $\mathbb{Z}_2$ in
Eqs.\eqref{g11} and \eqref{g11a} are
\begin{align}\nonumber
\mathbb{Z}_1&=\frac{2\Phi}{16\pi+\Phi\mathcal{R}}\bigg[\big(\frac{\varsigma
B\mathcal{R}^{10}}{A}\big)^.+\big(\frac{\varsigma
B\mathcal{R}^{11}}{A}\big)'-\big(\mu\mathcal{R}^{00}\big)^.-\big(\mu\mathcal{R}^{01}\big)'+\frac{2s^2\dot{C}\mathcal{G}^{00}}{C^5}\\\nonumber
&-\mathcal{G}^{01}\bigg(\frac{ss'}{C^4}-\frac{2s^2C'}{C^5}\bigg)+\frac{1}{2A^2}\bigg\{\mathcal{R}_{00}\bigg(\frac{\dot{\mu}}{A^2}
-\frac{2\mu\dot{A}}{A^3}\bigg)+2\bigg(\frac{\dot{\varsigma}}{AB}-\frac{\varsigma\dot{A}}{A}-\frac{\varsigma\dot{B}}{B}\bigg)\\\label{A5}
&\times\mathcal{R}_{01}+\mathcal{R}_{11}\bigg(\frac{\dot{P}_r}{B^2}
-\frac{2P_r\dot{B}}{B^3}\bigg)+2\mathcal{R}_{22}\bigg(\frac{\dot{P}_\bot}{C^2}-\frac{2P_\bot\dot{C}}{C^3}\bigg)\bigg\}
-\frac{\mu\dot{\mathcal{R}}}{2A^2}\bigg],\\\nonumber
\mathbb{Z}_2&=\frac{2\Phi}{16\pi+\Phi\mathcal{R}}\bigg[\big(P_r\mathcal{R}^{10}\big)^.+\big(P_r\mathcal{R}^{11}\big)'
-\big(\frac{\varsigma
A\mathcal{R}^{00}}{B}\big)^.-\big(\frac{\varsigma
A\mathcal{R}^{11}}{B}\big)'+\frac{2s^2\dot{C}\mathcal{G}^{10}}{C^5}\\\nonumber
&-\mathcal{G}^{11}\bigg(\frac{ss'}{C^4}-\frac{2s^2C'}{C^5}\bigg)-\frac{1}{2B^2}\bigg\{\mathcal{R}_{00}\bigg(\frac{\mu'}{A^2}-\frac{2\mu
A'}{A^3}\bigg)+2\bigg(\frac{\varsigma'}{AB}-\frac{\varsigma
A'}{A}-\frac{\varsigma B'}{B}\bigg)\\\label{A6}
&\times\mathcal{R}_{01}+\mathcal{R}_{11}\bigg(\frac{P'_r}{B^2}-\frac{2P_rB'}{B^3}\bigg)+2\mathcal{R}_{22}\bigg(\frac{P'_\bot}{C^2}
-\frac{2P_\bot
C'}{C^3}\bigg)\bigg\}-\frac{P_r\mathcal{R}'}{2B^2}\bigg].
\end{align}

\section*{Appendix B}

\renewcommand{\theequation}{B\arabic{equation}}
\setcounter{equation}{0} The scalars \eqref{g28}-\eqref{g31}
encompass modified corrections which are
\begin{align}\nonumber
\chi_{1}^{(\mathcal{C})}&=-\frac{8\pi\Phi}{1-\frac{\Phi
s^2}{2C^4}}\bigg[\bigg\{\Box\Omega^{\rho}_{\vartheta}
+\frac{1}{2}\nabla_{\varphi}\nabla^{\rho}\Omega^{\varphi}_{\vartheta}
+\frac{1}{2}\nabla_{\varphi}\nabla_{\vartheta}\Omega^{\varphi\rho}\bigg\}h^{\vartheta}
_{\rho}-\big(\mathcal{R}^{\rho}_{\varphi}h^{\varphi}_{\rho}\\\nonumber
&+\mathcal{R}_{\varphi\vartheta}h^{\vartheta\varphi}\big)\bigg(P-\frac{\Pi}{3}+\frac{s^2}{8\pi
C^4}\bigg)-\frac{\mathcal{Q}}{2}+\frac{1}{2}\nabla_{\varphi}\nabla_{\xi}\Omega^{\varphi\xi}-\frac{1}{2}\Box(\mu-3P)\\\label{B1}
&-\nabla_{\varphi}\nabla_{\rho}\Omega^{\varphi\rho}+2\mathcal{R}\bigg(P-\frac{\Pi}{3}+\frac{s^2}{8\pi
C^4}\bigg)-2g^{\rho\vartheta}\mathcal{R}^{\varphi\xi}\frac{\partial^2\mathbb{L}_{\mathcal{M}}}{\partial
g^{\rho\vartheta}\partial g^{\varphi\xi}}\bigg],\\\nonumber
\chi_{2}^{(\mathcal{C})}&=\frac{8\pi\Phi}{1-\frac{\Phi
s^2}{2C^4}}\bigg[\frac{1}{2}\nabla_{\varphi}\nabla_{\xi}\Omega^{\varphi\xi}\big\{\Box(\mu-3P)
-\mathcal{K}^{\varphi}\mathcal{K}^{\vartheta}\Box\Omega_{\varphi\vartheta}
+3\mathcal{K}_{\varphi}\mathcal{K}^{\vartheta}\Box\Omega^{\varphi}_{\vartheta}\big\}\\\nonumber
&+\frac{3\mathcal{Q}}{2}+4\mathcal{R}^{\vartheta}_{\varphi}\bigg\{\bigg(P+\frac{s^2}{24\pi
C^4}\bigg)h^{\varphi}_{\vartheta}+2\bigg(\Pi-\frac{s^2}{4\pi
C^4}\bigg)
\bigg(\mathcal{W}^{\varphi}\mathcal{W}_{\vartheta}-\frac{1}{3}h^{\varphi}_{\vartheta}\bigg)\\\nonumber
&+\varsigma\mathcal{K}^{\varphi}\mathcal{W}_{\vartheta}\bigg\}+6\mathcal{R}^{\vartheta}_{\varphi}\bigg\{\bigg(\mu-\frac{s^2}{8\pi
C^4}\bigg)\mathcal{K}^{\varphi}\mathcal{K}_{\vartheta}+\varsigma\mathcal{W}^{\varphi}\mathcal{K}_{\vartheta}\bigg\}
+6\mathcal{R}_{\varphi\vartheta}\\\nonumber
&\times\bigg\{\bigg(\mu-\frac{s^2}{8\pi
C^4}\bigg)\mathcal{K}^{\varphi}\mathcal{K}^{\vartheta}+\varsigma\mathcal{K}^{\vartheta}\mathcal{W}^{\varphi}\bigg\}
-2\nabla_{\varphi}\nabla^{\vartheta}\Omega^{\varphi}_{\vartheta}
+3\mathcal{K}_{\vartheta}\mathcal{K}^{\xi}\nabla_{\varphi}\nabla^{\vartheta}\Omega^{\varphi}_{\xi}\\\nonumber
&+3\mathcal{K}_{\vartheta}\mathcal{K}^{\xi}\nabla_{\varphi}\nabla_{\xi}\Omega^{\varphi\vartheta}
-2\mathcal{K}^{\vartheta}\mathcal{K}^{\xi}\nabla_{\varphi}\nabla_{\vartheta}\Omega^{\varphi}_{\xi}
-4h^{\vartheta\beta}\mathcal{R}^{\varphi\xi}\frac{\partial^2\mathbb{L}_{\mathcal{M}}}{\partial
g^{\vartheta\beta}\partial g^{\varphi\xi}}\\\label{B2}
&+\frac{1}{2}\Box(\mu-3P)+\nabla_{\varphi}\nabla_{\vartheta}
\Omega^{\varphi\vartheta}+2g^{\vartheta\xi}\mathcal{R}^{\alpha\rho}\frac{\partial^2\mathbb{L}_{\mathcal{M}}}{\partial
g^{\vartheta\xi}\partial g^{\alpha\rho}}\bigg],\\\nonumber
\chi_{\varphi\vartheta}^{(\mathcal{C})}&=-\frac{4\pi\Phi}{1-\frac{\Phi
s^2}{2C^4}}\bigg[-\frac{1}{2}\big\{h^{\lambda}_{\varphi}h^{\xi}_{\vartheta}\Box\Omega_{\lambda\xi}-\Box\Omega_{\varphi\vartheta}
-\mathcal{K}_{\varphi}\mathcal{K}_{\vartheta}\mathcal{K}_{\lambda}\mathcal{K}^{\delta}\Box\Omega^{\lambda}_{\delta}\big\}
-(h^{\lambda}_{\varphi}\mathcal{R}_{\lambda\rho}\\\nonumber
&-\mathcal{R}_{\varphi\rho})\bigg\{\bigg(P+\frac{s^2}{24\pi
C^4}\bigg)h^{\rho}_{\vartheta}+\bigg(\Pi-\frac{s^2}{4\pi
C^4}\bigg)\bigg(\mathcal{W}^{\rho}\mathcal{W}_{\vartheta}-\frac{1}{3}h^{\rho}_{\vartheta}\bigg)
+\varsigma\mathcal{K}^{\rho}\mathcal{W}_{\vartheta}\bigg\}\\\nonumber
&-(h^{\xi}_{\vartheta}\mathcal{R}_{\rho\xi}-\mathcal{R}_{\rho\vartheta})
\bigg\{\bigg(P+\frac{s^2}{24\pi C^4}\bigg)h^{\rho}_{\varphi}
+\bigg(\Pi-\frac{s^2}{4\pi
C^4}\bigg)\bigg(\mathcal{W}^{\rho}\mathcal{W}_{\varphi}-\frac{1}{3}h^{\rho}_{\varphi}\bigg)\\\nonumber
&+\varsigma\mathcal{K}^{\rho}\mathcal{W}_{\varphi}\bigg\}
+\frac{1}{2}\{h^{\lambda}_{\varphi}h^{\xi}_{\vartheta}\nabla_{\rho}\nabla_{\lambda}\Omega^{\rho}_{\xi}
+h^{\lambda}_{\varphi}h^{\xi}_{\vartheta}\nabla_{\rho}\nabla_{\xi}\Omega^{\rho}_{\lambda}
-\nabla_{\rho}\nabla_{\varphi}\Omega^{\rho}_{\vartheta}\\\nonumber
&-\nabla_{\rho}
\nabla_{\vartheta}\Omega^{\rho}_{\varphi}-\mathcal{K}_{\varphi}\mathcal{K}_{\vartheta}\mathcal{K}_{\lambda}\mathcal{K}^{\delta}
\nabla_{\rho}\nabla^{\lambda}\Omega^{\rho}_{\delta}
-\mathcal{K}_{\varphi}\mathcal{K}_{\vartheta}\mathcal{K}_{\lambda}\mathcal{K}^{\delta}\nabla_{\rho}\nabla_{\delta}\Omega^{\rho\lambda}\}
\\\label{B3}
&+2\mathcal{R}^{\rho\beta}h^{\lambda}_{\varphi}\bigg\{h^{\xi}_{\vartheta}\frac{\partial^2\mathbb{L}_{\mathcal{M}}}{\partial
g^{\lambda\xi}\partial g^{\rho\beta}}
-\frac{\partial^2\mathbb{L}_{\mathcal{M}}}{\partial
g^{\epsilon\vartheta}\partial g^{\rho\beta}}\bigg\}\bigg].
\end{align}

\noindent The complexity-free and homologous conditions in the
absence of dissipation flux are
\begin{align}\nonumber
&\epsilon-\frac{2\pi}{C^4}\left(\frac{\Phi\mathcal{R}}{2}+1\right)\big[\chi_9^{-1}\big\{\chi_7\big(2C^4-s^2\Phi\big)-\chi_4^{-1}\big(\chi_9
\big(\chi_1\big(s^2\Phi-2C^4\big)+2C^4s^2\\\nonumber &\times
\chi_6\big)+\chi_3\big(\chi_7\big(s^2
\Phi-2C^4\big)-2C^4s^2\chi_{11}\big)\big)+2C^4s^2\chi_{11}\big\}-2s^2-\big\{\chi_9\big(\big(\chi_5\chi_9\\\nonumber
&+\chi_3\big(\chi_{10}+8\pi\big)\big)\chi_{15}-\chi_4
\big(-8\pi\chi_{14}+\chi_{10}\big(\chi_{14}-8\pi\big)+\chi_9\chi_{16}+64\pi^2\big)\big)\big\}^{-1}\\\nonumber
&\times\big\{\big(8\pi-\chi_{10}\big)\big(\chi_{15}\big(\chi_9\big(\chi_1\big(s^2\Phi-2C^4\big)+2C^4s^2
\chi_6\big)+\chi_3\big(\chi_7\big(s^2\Phi-2C^4\big)\\\nonumber
&-2C^4s^2\chi_{11}\big)\big)+\chi_4
\big(\chi_7\big(8\pi-\chi_{14}\big)\big(s^2\Phi-2C^4\big)+\chi_9\big(\chi_{12}
\big(2C^4-s^2\Phi\big)+2C^4\\\nonumber &\times
s^2\chi_{17}\big)+2C^4s^2\chi_{11}\big(\chi_{14}-8\pi\big)\big)\big)\big\}-\big\{\chi_4\chi_9\big(\big(-\chi_5\chi_9
-\chi_3\big(\chi_{10}-8\pi\big)\big)\chi_{15}\\\nonumber
&+\chi_4\big(-8\pi\chi_{14}+\chi_{10}\big(\chi_{14}-8\pi\big)+\chi_9\chi_{16}+64\pi^2\big)\big)\big\}^{-1}
\big\{\big(\chi_5\chi_9+\chi _3 \big(\chi _{10}-8 \pi
\big)\big)\\\nonumber &\times \big(\chi _{15} \big(\chi _9 \big(\chi
_1 \big(s^2 \Phi -2 C^4\big)+2 C^4 s^2\chi _6\big)+\chi _3 \big(\chi
_7 \big(s^2 \Phi -2 C^4\big)-2 C^4 s^2 \chi
_{11}\big)\big)\\\nonumber &+\chi _4 \big(\chi _7 \big(8 \pi
-\chi_{14}\big) \big(s^2 \Phi -2 C^4\big)+\chi _9 \big(\chi _{12}
\big(2 C^4-s^2 \Phi \big)+2 C^4 s^2 \chi _{17}\big)+2 C^4
s^2\\\nonumber &\times \chi_{11} \big(\chi _{14}-8 \pi
\big)\big)\big)\big\} -\big\{\big(\big(-\chi _5 \chi _9-\chi _3
\big(\chi _{10}-8 \pi \big)\big) \chi _{15}+\chi _4 \big(-8 \pi \chi
_{14}+\chi _{10}\\\nonumber &\times \big(\chi _{14}-8 \pi \big)+\chi
_9 \chi _{16}+64 \pi ^2\big)\big) \big(\big(\big(-\chi _5 \chi
_9-\chi _3 \big(\chi _{10}-8 \pi \big)\big) \chi _{15}-\chi _4
\big(8 \pi \chi _{14}\\\nonumber &-\chi _{10} \big(\chi _{14}-8 \pi
\big)-\chi _9 \chi _{16}-64 \pi ^2\big)\big) \big(-\chi _4 \chi _8
\chi _{20}+\chi _3 \chi _8 \big(\chi _{21}-8 \pi \big)+\chi _9
\big(-\chi _4\\\nonumber &\times \chi _{19}-8 \pi \chi _{21}+\chi _2
\big(\chi _{21}-8 \pi \big)+64 \pi^2\big)\big)+\big(\big(\chi _3
\chi _8+\big(\chi _2-8 \pi \big) \chi _9\big) \chi _{15}-\chi _4
\\\nonumber &\times\big(\chi _9 \chi _{13}+\chi _8 \big(\chi _{14}-8 \pi
\big)\big)\big) \big(\chi _3 \big(8 \pi -\chi _{10}\big) \big(8 \pi
-\chi _{21}\big)+\chi _5 \chi _9 \big(\chi _{21}-8 \pi \big)+\chi _4
\\\nonumber &\times\big(\big(8 \pi -\chi _{10}\big) \chi _{20}-\chi _9 \chi
_{22}\big)\big)\big)\big\}^{-1}\big\{\big(\big(\chi _5 \chi _8+\chi
_2 \big(8 \pi -\chi _{10}\big)+8 \pi \big(\chi _{10}-8 \pi
\big)\big) \chi _{15}\\\nonumber &+\chi _4 \big(\big(\chi _{10}-8
\pi \big) \chi _{13}-\chi _8 \chi _{16}\big)\big) \big(\big(\chi _3
\big(8 \pi -\chi _{10}\big) \big(8 \pi -\chi _{21}\big)+\chi _5 \chi
_9 \big(\chi _{21}-8 \pi \big)\\\nonumber &+\chi _4 \big(\big(8 \pi
-\chi _{10}\big) \chi _{20}-\chi _9 \chi _{22}\big)\big) \big(\chi
_{15} \big(\chi _9 \big(\chi _1 \big(s^2 \Phi -2 C^4\big)+2 C^4 s^2
\chi _6\big)+\chi _3 \big(\chi _7 \\\nonumber &\times\big(s^2 \Phi
-2 C^4\big)-2 C^4 s^2 \chi _{11}\big)\big)-\chi _4 \big(\chi _7
\big(8 \pi -\chi _{14}\big) \big(2 C^4-s^2 \Phi \big)+\chi _9
\big(\chi _{12} \big(s^2 \Phi\\\nonumber &-2 C^4\big)-2 C^4 s^2 \chi
_{17}\big)+2 C^4 s^2 \chi _{11} \big(8 \pi -\chi
_{14}\big)\big)\big)+\big(\big(-\chi _5 \chi _9-\chi _3 \big(\chi
_{10}-8 \pi \big)\big)\\\nonumber &\times \chi _{15}+\chi _4 \big(-8
\pi \chi _{14}+\chi _{10} \big(\chi _{14}-8 \pi \big)+\chi _9 \chi
_{16}+64 \pi^2\big)\big) \big(\chi _1 \chi _9 \big(8 \pi -\chi
_{21}\big)\\\nonumber &\times \big(2 C^4-s^2 \Phi \big)+\chi _3
\big(8 \pi -\chi _{21}\big) \big(\chi _7 \big(2 C^4-s^2 \Phi \big)+2
C^4 s^2 \chi _{11}\big)-16 \pi  C^4 s^2 \chi _6 \chi _9\\\nonumber
&+2 C^4 s^2 \chi _4 \chi _{11} \chi _{20}+2 C^4 s^2 \chi _6 \chi _9
\chi _{21}-2 C^4 s^2 \chi _4 \chi _9 \chi _{23}+2 C^4 \chi _4 \chi
_9 \chi _{18}+2 C^4 \chi _4 \\\nonumber &\times\chi _7 \chi
_{20}-s^2 \Phi \chi _4 \chi _9 \chi _{18}-s^2 \Phi \chi _4 \chi _7
\chi_{20}\big)\big)\big\}+\big\{\big(\big(-\chi _5 \chi _9-\chi _3
\big(\chi _{10}-8 \pi \big)\big) \chi _{15}\\\nonumber &+\chi _4
\big(-8 \pi \chi _{14}+\chi _{10} \big(\chi _{14}-8 \pi \big)+\chi
_9 \chi _{16}+64 \pi ^2\big)\big) \big(\big(\big(-\chi _5 \chi _9-
\big(\chi _{10}-8 \pi \big)\\\nonumber &\times\chi _3\big) \chi
_{15}+\chi _4 \big(-8 \pi \chi _{14}+\chi _{10} \big(\chi _{14}-8
\pi \big)+\chi _9 \chi _{16}+64 \pi ^2\big)\big) \big(-\chi _4 \chi
_8 \chi _{20}\\\nonumber &+\chi _3 \chi _8 \big(\chi _{21}-8 \pi
\big)+\chi _9 \big(-\chi _4 \chi _{19}-8 \pi \chi _{21}+\chi _2
\big(\chi _{21}-8 \pi \big)+64 \pi ^2\big)\big)+\big(\big(\chi _3
\\\nonumber &\times\chi _8+\big(\chi _2-8 \pi \big) \chi _9\big) \chi
_{15}-\chi _4 \big(\chi _9 \chi _{13}+\chi _8 \big(\chi _{14}-8 \pi
\big)\big)\big) \big(\chi _3 \big(8 \pi -\chi _{10}\big) \big(8 \pi
\\\nonumber &-\chi _{21}\big)+\chi _5 \chi _9 \big(\chi _{21}-8 \pi \big)+\chi
_4 \big(\big(8 \pi -\chi _{10}\big) \chi _{20}-\chi _9 \chi
_{22}\big)\big)\big)\big\}^{-1}\big\{\big(-\chi _3 \chi _{13} \chi
_{10}\\\nonumber &-8 \pi  \chi _{14} \chi _{10}+64 \pi ^2 \chi
_{10}+8 \pi \chi _3 \chi _{13}+64 \pi ^2 \chi _{14}-\chi _5
\big(\chi _9 \chi _{13}+\chi _8 \big(\chi _{14}-8 \pi
\big)\big)\\\nonumber &+\chi _3 \chi _8 \chi _{16}-8 \pi  \chi _9
\chi _{16}-512 \pi ^3+ \big(-8 \pi  \chi _{14}+\chi _{10} \big(\chi
_{14}-8 \pi \big)+\chi _9 \chi _{16}+64 \pi ^2\big)\\\nonumber
&\times\chi _2\big) \big(\big(\chi _3 \big(8 \pi -\chi _{10}\big)
\big(8 \pi -\chi _{21}\big)+\chi _5 \chi _9 \big(\chi _{21}-8 \pi
\big)+\chi _4 \big(\big(8 \pi -\chi _{10}\big) \chi _{20}-\chi _9
\\\nonumber &\times\chi _{22}\big)\big) \big(\chi _{15} \big(\chi _9 \big(\chi _1
\big(s^2 \Phi -2 C^4\big)+2 C^4 s^2 \chi _6\big)+ \big(\chi _7
\big(s^2 \Phi -2 C^4\big)-2 C^4 s^2 \chi _{11}\big)\\\nonumber
&\times\chi _3\big)-\chi _4 \big(\chi _7 \big(8 \pi -\chi _{14}\big)
\big(2 C^4-s^2 \Phi \big)+\chi _9 \big(\chi _{12} \big(s^2 \Phi -2
C^4\big)-2 C^4 s^2 \chi _{17}\big)\\\nonumber &+2 C^4 s^2 \chi _{11}
\big(8 \pi -\chi _{14}\big)\big)\big)+\big(\big(-\chi _5 \chi
_9-\chi _3 \big(\chi _{10}-8 \pi \big)\big) \chi _{15}+\chi _4
\big(-8 \pi \chi _{14}\\\nonumber &+\chi _{10} \big(\chi _{14}-8 \pi
\big)+\chi _9 \chi _{16}+64 \pi ^2\big)\big) \big(\chi _1 \chi _9
\big(8 \pi -\chi _{21}\big) \big(2 C^4-s^2 \Phi \big)+\chi _3 \big(8
\pi \\\nonumber &-\chi _{21}\big) \big(\chi _7 \big(2 C^4-s^2 \Phi
\big)+2 C^4 s^2 \chi _{11}\big)-16 \pi  C^4 s^2 \chi _6 \chi _9+2
C^4 s^2 \chi _4 \chi _{11} \chi _{20}+2 C^4\\\nonumber &\times s^2
\chi _6 \chi _9 \chi _{21}-2 C^4 s^2 \chi _4 \chi _9 \chi _{23}+2
C^4 \chi _4 \chi _9 \chi _{18}+2 C^4 \chi _4 \chi _7 \chi _{20}-s^2
\Phi \chi _4 \chi _9 \chi _{18}\\\label{g51} &-s^2 \Phi \chi _4 \chi
_7 \chi _{20}\big)\big)\big\}\big]=0,\\\nonumber
&\big(\chi_4\big(8\pi-\chi_{14}\big)+\chi_3\chi_{15}\big)\big(-\chi_7\big(2C^4-s^2\Phi\big)-2
C^4s^2\chi_{11}\big)+\chi_9\big(\chi_{15}\big(\chi_1\big(s^2\Phi\\\nonumber
&-2C^4\big)+2C^4s^2\chi_6\big)+\chi_4
\big(\chi_{12}\big(2C^4-s^2\Phi\big)+2C^4s^2\chi_{17}\big)\big)+\big\{\big(\big(-\chi_5\chi_9-\chi_3\\\nonumber
&\times\big(\chi_{10}-8\pi\big)\big)\chi_{15}+\chi_4\big(-8\pi\chi_{14}+\chi_{10}\big(\chi_{14}-8\pi
\big)+\chi_9\chi_{16}+64\pi^2\big)\big)\big(-\chi_4\\\nonumber
&\times\chi_8\chi_{20}+\chi_3\chi_8
\big(\chi_{21}-8\pi\big)+\chi_9\big(-\chi_4\chi_{19}-8\pi\chi_{21}+\chi_2
\big(\chi_{21}-8\pi\big)+64\pi^2\big)\big)\\\nonumber
&+\big(\big(\chi_3\chi_8+\big(\chi_2-8\pi\big)
\chi_9\big)\chi_{15}-\chi_4\big(\chi_9\chi_{13}+\chi_8\big(\chi_{14}-8\pi\big)\big)\big)\big(\chi_3
\big(8\pi-\chi_{10}\big)\\\nonumber
&\times\big(8\pi-\chi_{21}\big)+\chi_5\chi_9\big(\chi
_{21}-8\pi\big)+\chi_4\big(\big(8\pi-\chi_{10}\big)\chi_{20}-\chi_9\chi_{22}\big)\big)\big\}^{-1}
\big\{\big(\chi _4 \big(\chi _9\\\nonumber &\times\chi _{13}+\chi _8
\big(\chi _{14}-8 \pi \big)\big)-\big(\chi _3 \chi _8+\big(\chi _2-8
\pi \big) \chi _9\big) \chi _{15}\big) \big(\big( \big(8 \pi -\chi
_{10}\big) \big(8 \pi -\chi _{21}\big)\\\nonumber &\times\chi
_3+\chi _5 \chi _9 \big(\chi _{21}-8 \pi \big)+\chi _4 \big(\big(8
\pi -\chi _{10}\big) \chi _{20}-\chi _9 \chi _{22}\big)\big)
\big(\chi _{15} \big(-\big(\chi _9 \big(\chi _1 \big(2
C^4\\\nonumber &-s^2 \Phi \big)-2 C^4 s^2 \chi _6\big)+\chi _3
\big(\chi _7 \big(2 C^4-s^2 \Phi \big)+2 C^4 s^2 \chi
_{11}\big)\big)\big)-\chi _4 \big(\chi _7 \big(8 \pi -\chi
_{14}\big)\\\nonumber &\times\big(2 C^4-s^2 \Phi \big)+\chi _9
\big(\chi _{12} \big(s^2 \Phi -2 C^4\big)-2 C^4 s^2 \chi
_{17}\big)+2 C^4 s^2 \chi _{11} \big(8 \pi -\chi
_{14}\big)\big)\big)\\\nonumber &+\big(\big(-\chi _5 \chi _9-\chi _3
\big(\chi _{10}-8 \pi \big)\big) \chi _{15}+\chi _4 \big(-8 \pi \chi
_{14}+\chi _{10} \big(\chi _{14}-8 \pi \big)+\chi _9 \chi
_{16}\\\nonumber &+64 \pi ^2\big)\big) \big(\chi _1 \chi _9 \big(8
\pi -\chi _{21}\big) \big(2 C^4-s^2 \Phi \big)+\chi _3 \big(8 \pi
-\chi _{21}\big) \big(\chi _7 \big(2 C^4-s^2 \Phi \big)+2\\\nonumber
&\times C^4 s^2 \chi _{11}\big)-16 \pi  C^4 s^2 \chi _6 \chi _9+2
C^4 s^2 \chi _4 \chi _{11} \chi _{20}+2 C^4 s^2 \chi _6 \chi _9 \chi
_{21}-2 C^4 s^2 \chi _4 \chi _9\\\label{g52} &\times \chi _{23}+2
C^4 \chi _4 \chi _9 \chi _{18}+2 C^4 \chi _4 \chi _7 \chi _{20}-s^2
\Phi \chi _4 \chi _9 \chi _{18}-s^2 \Phi \chi _4 \chi _7 \chi
_{20}\big)\big)\big\}=0.
\end{align}

\noindent The homologous condition in the presence of heat
dissipation is
\begin{align}\nonumber
\bar{\varsigma}&=\frac{8\pi}{8\pi-\chi_{10}}\bigg[\frac{\chi_7}{8\pi}\bigg(1-\frac{s^2\Phi}{2C^4}\bigg)+\frac{s^2\chi_{11}}{8\pi}
+\big\{16 \pi  C^4 \big(\big(\big(-\chi _5 \chi _9-\chi _3 \big(\chi
_{10}-8 \pi \big)\big)\\\nonumber &\times\chi _{15}+\chi _4 \big(-8
\pi \chi _{14}+\chi _{10} \big(\chi _{14}-8 \pi \big)+\chi _9 \chi
_{16}+64 \pi ^2\big)\big) \big(-\chi _4 \chi _8 \chi _{20}+\chi _3
\chi _8\\\nonumber &\times \big(\chi _{21}-8 \pi \big)+\chi _9
\big(-\chi _4 \chi _{19}-8 \pi \chi _{21}+\chi _2 \big(\chi _{21}-8
\pi \big)+64 \pi ^2\big)\big)+\big(\big(\chi _3 \chi_8\\\nonumber
&+\big(\chi _2-8 \pi \big) \chi _9\big) \chi _{15}-\chi _4 \big(\chi
_9 \chi _{13}+\chi _8 \big(\chi _{14}-8 \pi \big)\big)\big)
\big(\chi _3 \big(8 \pi -\chi _{10}\big) \big(8 \pi -\chi
_{21}\big)\\\nonumber &+\chi _5 \chi _9 \big(\chi _{21}-8 \pi
\big)+\chi _4 \big(\big(8 \pi -\chi _{10}\big) \chi _{20}-\chi _9
\chi _{22}\big)\big)\big)\big\}^{-1}\big\{\chi _8 \big(\big(\chi _3
\big(8 \pi -\chi _{10}\big) \\\nonumber &\times\big(8 \pi -\chi
_{21}\big)+\chi _5 \chi _9 \big(\chi _{21}-8 \pi \big)+\chi _4
\big(\big(8 \pi -\chi _{10}\big) \chi _{20}-\chi _9 \chi
_{22}\big)\big) \big(\chi _{15} \big(\chi _9 \big(2 s^2\\\nonumber
&\times C^4 \chi _6-\chi _1 \big(2 C^4-s^2 \Phi \big)\big)-\chi _3
\big(\chi _7 \big(2 C^4-s^2 \Phi \big)+2 C^4 s^2 \chi
_{11}\big)\big)-\chi _4 \big(\chi _7 \big(8 \pi\\\nonumber & -\chi
_{14}\big) \big(2 C^4-s^2 \Phi \big)+\chi _9 \big(\chi _{12}
\big(s^2 \Phi -2 C^4\big)-2 C^4 s^2 \chi _{17}\big)+2 C^4 \big(8 \pi
-\chi _{14}\big)\\\nonumber &\times s^2 \chi
_{11}\big)\big)+\big(\big(-\chi _5 \chi _9-\chi _3 \big(\chi _{10}-8
\pi \big)\big) \chi _{15}+\chi _4 \big(-8 \pi \chi _{14}+\chi _{10}
\big(\chi _{14}-8 \pi \big)\\\nonumber &+\chi _9 \chi _{16}+64 \pi
^2\big)\big) \big(\chi _1 \chi _9 \big(8 \pi -\chi _{21}\big) \big(2
C^4-s^2 \Phi \big)+\chi _3 \big( \big(2 C^4-s^2 \Phi \big)\chi _7+2
C^4\\\nonumber &\times s^2 \chi _{11}\big)\big(8 \pi -\chi
_{21}\big)-16 \pi C^4 s^2 \chi _6 \chi _9+2 C^4 s^2 \chi _4 \chi
_{11} \chi _{20}+2 C^4 s^2 \chi _6 \chi _9 \chi _{21}-2
s^2\\\nonumber & \times C^4 \chi _4 \chi _9 \chi _{23}+2 C^4 \chi _4
\chi _9 \chi _{18}+2 C^4 \chi _4 \chi _7 \chi _{20}-s^2 \Phi \chi _4
\chi _9 \chi _{18}-s^2 \Phi \chi _4 \chi _7 \chi
_{20}\big)\big)\big\}\bigg]\\\label{g53} &-\frac{\sigma C'}{4\pi
BC}.
\end{align}

\noindent The values of $\chi_i$, $i=1,2,3,...,23$ used in
Eqs.\eqref{g51}-\eqref{g53} are
\begin{align}\label{B4}
\chi_1&=-\frac{1}{B^2}\bigg(\frac{2C''}{C}-\frac{2B'C'}{BC}-\frac{B^2}{C^2}+\frac{C'^2}{C^2}\bigg)
+\frac{\dot{C}}{C}\bigg(\frac{2\dot{B}}{B}+\frac{\dot{C}}{C}\bigg),\\\label{B5}
\chi_2&=\Phi\bigg(\frac{2\dot{B}\dot{C}}{BC}-\frac{3\ddot{B}}{2B^2}+\frac{\dot{C}^2}{C^2}-\frac{3\ddot{C}}{C}
+\frac{\mathcal{R}}{2}\bigg),\\\label{B6}
\chi_3&=\Phi\bigg(\frac{2B'\dot{C}}{B^3C}-\frac{2B'C'}{B^3C}+\frac{\dot{B}\dot{C}}{BC}+\frac{\ddot{B}}{2B}+\frac{C'^2}{B^2
C^2}+\frac{C''}{B^2C}\bigg),\\\label{B7}
\chi_4&=\Phi\bigg(\frac{B'C'}{B^3C}+\frac{\dot{B}\dot{C}}{BC}-\frac{C'^2}{B^2C^2}-\frac{C''}{B^2C}+\frac{\dot{C}^2}{C}
+\frac{\ddot{C}}{C}\bigg),\\\label{B8}
\chi_5&=\Phi\bigg(\frac{5\dot{B}C'}{BC^2}+\frac{2C'\dot{C}}{BC^2}-\frac{\dot{C}'}{BC}\bigg),\\\label{B9}
\chi_6&=\frac{1}{C^4}+\frac{\Phi}{B^3C^5}\bigg(2\dot{B}B^2\dot{C}+2B'C'+\ddot{B}B^2C-2BC''-\frac{\mathcal{R}B^3C}{2}\bigg),\\\label{B10}
\chi_7&=-\frac{1}{B}\bigg(\frac{2\dot{B}C'}{BC}-\frac{2\dot{C}'}{C}\bigg),
\quad
\chi_8=\chi_9=\Phi\bigg(\frac{2\dot{C}'}{BC}-\frac{2\dot{B}C'}{B^2C}\bigg),\\\label{B11}
\chi_{10}&=\Phi\bigg(\frac{2C''}{B^2C}-\frac{2B'C'}{B^3C}-\frac{\ddot{B}}{B}-\frac{2\ddot{C}}{C}+\frac{\mathcal{R}}{2}\bigg),\\\label{B12}
\chi_{11}&=\frac{\Phi}{B^2}\left(\dot{B}C'-B\dot{C}'\right), \quad
\chi_{12}=\frac{C'^2}{B^2C^2}-\frac{2\ddot{C}}{C}-\frac{\dot{C}^2}{C^2}-\frac{1}{C^2},\\\label{B13}
\chi_{13}&=\Phi\bigg(\frac{\ddot{B}}{2B}-\frac{\ddot{C}}{C}-\frac{\dot{C}^2}{C^2}\bigg),\\\label{B14}
\chi_{14}&=\Phi\bigg(\frac{3C''}{B^2C}-\frac{2B'C'}{B^3C}-\frac{2B'\dot{C}}{B^3C}-\frac{3\dot{B}\dot{C}}{BC}-\frac{3
\ddot{B}}{2B}-\frac{C'^2}{B^2C^2}+\frac{\mathcal{R}}{2}\bigg),\\\label{B15}
\chi_{15}&=\Phi\bigg(\frac{C'^2}{B^2C^2}-\frac{B'C'}{B^3C}-\frac{\dot{B}\dot{C}}{BC}+\frac{C''}{B^2C}-\frac{\dot{C}^2}{C^2}
-\frac{\ddot{C}}{C}\bigg),\\\label{B16}
\chi_{16}&=\Phi\bigg(\frac{2\dot{C}'}{BC}-\frac{4\dot{B}C'}{B^2C}-\frac{2C'\dot{C}}{BC^2}\bigg),\\\label{B17}
\chi_{17}&=\frac{1}{C^4}-\frac{\Phi}{B^3C^5}\left(\ddot{B}B^2C+2B^3\ddot{C}-\frac{\mathcal{R}B^3C}{2}\right),\\\label{B18}
\chi_{18}&=\frac{1}{B^2}\bigg(\frac{C''}{C}-\frac{B'C'}{BC}\bigg)-\frac{\dot{B}\dot{C}}{BC}-\frac{\ddot{B}}{B}-\frac{\ddot{C}}{C},\\\label{B19}
\chi_{19}&=-\Phi\bigg(\frac{\dot{B}\dot{C}}{BC}+\frac{\ddot{B}}{2B}\bigg),
\quad
\chi_{20}=\Phi\bigg(\frac{B'C'}{B^3C}-\frac{2B'\dot{C}}{B^3C}-\frac{\dot{B}\dot{C}}{BC}-\frac{\ddot{B}}{2B}\bigg),\\\label{B20}
\chi_{21}&=\Phi\bigg(\frac{2C'^2}{B^2C^2}-\frac{2B'C'}{B^3C}-\frac{2\dot{B}\dot{C}}{BC}+\frac{2C''}{B^2C}-\frac{2\dot{C}^2}{C^2}
-\frac{2\ddot{C}}{C}-\frac{2}{C^2}+\frac{\mathcal{R}}{2}\bigg),\\\label{B21}
\chi_{22}&=\Phi\bigg(\frac{B'\dot{B}}{B^3}-\frac{2\dot{B}C'}{B^2C}-\frac{\dot{B}'}{B^2}\bigg),
\quad
\chi_{23}=\frac{1}{C^4}\bigg(1+\frac{\Phi\mathcal{R}}{2}\bigg).
\end{align}
\textbf{Data Availability Statement:} This manuscript has no
associated data.


\begin{thebibliography}{43}

\bibitem{1a} Perlmutter, S. et al.: Nature \textbf{391}(1998)51.

\bibitem{1b} Riess, A.G. et al.: Astron. J. \textbf{116}(1998)1009.

\bibitem{1c} Buchdahl, H.A.: Mon. Not. R. Astron. Soc. \textbf{150}(1970)1.

\bibitem{2} Nojiri, S. and Odintsov, S.D.: Phys. Rev. D \textbf{68}(2003)123512.

\bibitem{2a} Cognola, G. et al.: J. Cosmol. Astropart. Phys. \textbf{2005}(2005)010.

\bibitem{2b} Song, Y.S., Hu, W. and Sawicki, I.: Phys. Rev. D \textbf{75}(2007)044004.

\bibitem{2d} Sharif, M. and Yousaf, Z.: Astrophys. Space Sci. \textbf{351}(2014)351.

\bibitem{10} Bertolami, O. et al.: Phys. Rev. D \textbf{75}(2007)104016.

\bibitem{20} Harko, T. et al.: Phys. Rev. D \textbf{84}(2011)024020.

\bibitem{21} Sharif, M. and Zubair, M.: J. Exp. Theor. Phys. \textbf{117}(2013)248.

\bibitem{21a} Shabani, H. and Farhoudi, M.: Phys. Rev. D \textbf{88}(2013)044048.

\bibitem{21d} Moraes, P.H.R.S., Arba{\~n}il, J.D.V. and Malheiro, M.: J. Cosmol. Astropart. Phys. \textbf{2016}(2016)005.

\bibitem{21e} Sharif, M. and Siddiqa, A.: Eur. Phys. J. Plus \textbf{132}(2017)1.

\bibitem{21f} Das, A. et al.: Phys. Rev. D \textbf{95}(2017)124011.

\bibitem{22} Haghani, Z. et al.: Phys. Rev. D \textbf{88}(2013)044023.

\bibitem{22a} Sharif, M. and Zubair, M.: J. Cosmol. Astropart. Phys. \textbf{2013}(2013)042.

\bibitem{22b} Sharif, M. and Zubair, M.: J. High Energy Phys. \textbf{2013}(2013)79.

\bibitem{23} Odintsov, S.D. and S{\'a}ez-G{\'o}mez, D.: Phys. Lett. B \textbf{725}(2013)437.

\bibitem{25a} Sharif, M. and Waseem, A.: Can. J. Phys. \textbf{94}(2016)1024.

\bibitem{26} Yousaf, Z., Bhatti, M.Z. and Naseer, T.: Eur. Phys. J. Plus
\textbf{135}(2020)353; Yousaf, Z. et al.: Mon. Not. R. Astron. Soc.
\textbf{495}(2020)4334.

\bibitem{26a} Yousaf, Z., Bhatti, M.Z. and Naseer, T.: Ann. Phys. \textbf{420}(2020)168267; Int. J. Mod. Phys. D
\textbf{29}(2020)2050061.

\bibitem{26d} Yousaf, Z., Bhatti, M.Z. and Naseer, T.: Phys. Dark Universe \textbf{28}(2020)100535;
Yousaf, Z. et al.: Phys. Dark Universe \textbf{29}(2020)100581.

\bibitem{26e} Sharif, M. and Naseer, T.: Chin. J. Phys. \textbf{77}(2022)2655.

\bibitem{27} Sharif, M. and Naseer, T.: Chin. J. Phys.
\textbf{73}(2021)179; Phys. Scr. \textbf{97}(2022)055004; Pramana
\textbf{96}(2022)119; Indian J. Phys. (2022)1.

\bibitem{27aa} Naseer, T. and Sharif, M.: Universe \textbf{8}(2022)62.

\bibitem{35d} Bronnikov, K.A. and Kovalchuk, M.A.: Probl. teor. gravit. \`{e}lem.
\textbf{11}(1980)131.

\bibitem{35e} Wang, A.: Phys. Rev. D \textbf{68}(2003)064006.

\bibitem{27a} Bekenstein, J.D.: Phys. Rev. D \textbf{4}(1971)2185.

\bibitem{27b} Esculpi, M. and Aloma, E.: Eur. Phys. J. C \textbf{67}(2010)521.

\bibitem{27c} Sharif, M. and Azam, M.: Mon. Not. R. Astron. Soc. \textbf{430}(2013)3048.

\bibitem{27d} Takisa, P.M. and Maharaj, S.D.: Gen. Relativ. Gravit. \textbf{45}(2013)1951.

\bibitem{28} L\'{o}pez-Ruiz, R., Mancini, H.L. and Calbet, X.: Phys. Lett. A \textbf{209}(1995)321.

\bibitem{29} Calbet, X. and L{\'o}pez-Ruiz, R.: Phys. Rev. E \textbf{63}(2001)066116.

\bibitem{29a} Catal{\'a}n, R.G., Garay, J. and L{\'o}pez-Ruiz, R.: Phys. Rev. E \textbf{66}(2002)011102.

\bibitem{30} Sa{\~n}udo, J. and L{\'o}pez-Ruiz, R.: Phys. Lett. A \textbf{372}(2008)5283.

\bibitem{30a} Sa{\~n}udo, J. and Pacheco, A.F.: Phys. Lett. A \textbf{373}(2009)807.

\bibitem{31} Herrera, L.: Phys. Rev. D \textbf{97}(2018)044010.

\bibitem{32} Sharif, M. and Butt, I.I.: Eur. Phys. J. C \textbf{78}(2018)688.

\bibitem{32a} Sharif, M. and Butt, I.I.: Eur. Phys. J. C \textbf{78}(2018)850.

\bibitem{33} Herrera, L., Di Prisco, A. and Ospino, J.: Phys. Rev. D \textbf{98}(2018)104059.

\bibitem{34} Herrera, L., Di Prisco, A. and Ospino, J.: Phys. Rev. D \textbf{99}(2019)044049.

\bibitem{35} Sharif, M. and Majid, A.: Chin. J. Phys. \textbf{61}(2019)38.

\bibitem{35a} Sharif, M. and Majid, A.: Eur. Phys. J. C \textbf{80}(2020)1.

\bibitem{35c} Sharif, M. and Majid, A.: Indian J. Phys. \textbf{95}(2021)769.

\bibitem{35ca} Zubair, M. and Azmat, H.: Int. J. Mod. Phys. D \textbf{29}(2020)2050014.

\bibitem{36a} Zubair, M. and Azmat, H.: Phys. Dark Universe \textbf{28}(2020)100531.

\bibitem{36b} Sharif, M. and Hassan, K.: Chin. J. Phys. \textbf{77}(2022)1479; Mod. Phys. Lett. A \textbf{37}(2022)2250027.

\bibitem{36c} Herrera, L. et al.: Int. J. Mod. Phys. D \textbf{14}(2005)657.

\bibitem{36d} Herrera, L., Di Prisco, A. and Ospino, J.: Gen. Relativ. Gravit.
\textbf{44}(2012)2645.

\bibitem{41ba} Thorne, K.S.: Phys. Rev. \textbf{139}(1965)B244.

\bibitem{41bb} Herrera, L. et al.: Phys. Rev. D \textbf{79}(2009)064025.

\bibitem{42bc} Kippenhahn, R. and Weigert, A.: \emph{Stellar Structure and Evolution} (Springer, 1990).

\bibitem{42bd} Hansen, C.J., Kawaler, S.D. and Trimble, V.: \emph{Stellar Interiors: Physical Principles, Structure and Evolution} (Springer,
1994).
\end{thebibliography}
\end{document}